\documentclass{aa}
\usepackage{graphics}
\usepackage{psfig}
\usepackage{times}
\newcommand{\msun}{$M_\odot$}

\def\h2{H{\small II}}
\begin{document}
% -------------------------------------------------------------------
  \thesaurus{  03          % Extragalactic Astronomy
              (11.01.1;    % Galaxies: abundances
               11.04.2;    % Galaxies: dwarf
               11.05.2;    % Galaxies: evolution
               11.03.2;    % Galaxies: compact
               11.19.3;    % Galaxies: starburst
               11.19.5; )} % Galaxies: stellar content

\title{The Cometary Blue Compact Dwarf Galaxies \object{Mkn\,59} and \object{Mkn\,71}:}
\subtitle{A Clue to Dwarf Galaxy Evolution?}

\author{K.G.\ Noeske\inst{1}
\and N.G.\ Guseva \inst{2}
\and K.J.\ Fricke   \inst{1}
\and Y.I.\ Izotov \inst{2}
\and P.\ Papaderos \inst{1}
\and T.X.\ Thuan\inst{3}}
\offprints{knoeske@uni-sw.gwdg.de}
\institute{   Universit\"ats--Sternwarte, Geismarlandstra\ss e 11,
                 D--37083 G\"ottingen, Germany
\and
                 Main Astronomical Observatory
                 of National Academy of Sciences of Ukraine,
                 Goloseevo, 252650 Kiev-22, Ukraine
\and
                 Astronomy Department, University of Virginia, Charlottesville, 
		 VA 22903, USA
}
\date{Received(00.00.0000)/Accepted(00.00.0000)}
\titlerunning{Cometary BCDs}
\authorrunning{Noeske et al.}
\maketitle
% ======================================================================================
\begin{abstract}
``Cometary'' Blue Compact Dwarf Galaxies (iI,C BCDs) are characterized by an 
off--center starburst close to the end of their elongated stellar bodies. 
This rare phenomenon may carry some clues on how collective star formation
ignites and propagates in gas--rich low--mass stellar systems.
This off--center burst may be a fortuitous enhancement of the otherwise 
moderate star--forming activity of a dwarf irregular (dI), or may be caused 
by a set of special properties of such systems or their environment. 
We attempt here a first investigation of this issue by analysing two prototypical 
examples of cometary dwarf galaxies, the nearby iI,C BCDs \object{Markarian\,59} and 
\object{Markarian\,71}, both containing an extraordinarily luminous \ion{H}{ii} 
region in the outskirts of a dI--like host.
Using deep ground--based spectrophotometric data
\footnote{Obtained at the German--Spanish Astronomical Center, Calar Alto, operated by the 
Max--Planck--Institute for Astronomy, Heidelberg, jointly with the Spanish National Commission 
for Astronomy.}$^,$\footnote{Obtained at the Kitt Peak National Observatory, National Optical 
Astronomy Observatories, operated by the Association of Universities for Research in Astronomy, 
Inc., under cooperative agreement with the National Science Foundation.}
 and HST images
\footnote{Based on observations with the NASA/ESA \emph{Hubble Space Telescope}, 
obtained at the Space Telescope Science Institute, which is operated by AURA, INC., under
NASA contract No. NAS 5--26555.},
we study the physical state of the starburst regions and the structural properties of the 
underlying irregular galaxies. 
We find that the average metallicities show small scatter in the vicinity of the star-forming 
regions and along the major axis of \object{Mkn\,59} which suggests that mixing of heavy elements 
must have been efficient on scales of several kpc.  
The azimuthally averaged radial intensity distributions of the underlying host galaxies in either
 iI,C BCD can be approximated by an exponential law with a central surface brightness and scale 
length that is intermediate between typical iE/nE BCDs and dwarf irregulars.
Spectral population synthesis models in combination with colour magnitude diagrams and
colour profiles yield a most probable formation age of $\sim$ 2 Gyr for the low surface 
brightness (LSB) host galaxies in both iI,C BCDs, with upper age limits of $\sim$ 4 Gyr for 
\object{Mkn\,59} and $\sim$ 3 Gyr for \object{Mkn\,71}, i.e.
significantly lower than the typical age of several Gyr derived for the LSB component of 
iE/nE BCDs. 
These findings raise the question whether iI,C systems form a distinct physical 
class within BCDs with respect to the age and structural properties of their hosts, or 
whether they represent an evolutionary stage connecting young i0 BCDs and ``classical'' iE/nE BCDs.
In spite of the scarcity of available data, a review of the properties of analogous objects 
studied in the local universe and at medium redshifts provides some support for this
evolutionary hypothesis.
\end{abstract}

%---------------------------------------------------------------------------------------------


%----------------------------------------------------------------------------------------------

\section {Introduction}
\label{intro}

Despite the effort that has been devoted to the investigation of Blue Compact 
Dwarf Galaxies (BCDs), the origin of their recurrent starburst activity 
as well as its impact on their spectrophotometric and dynamical evolution
is still poorly understood.
A thorough investigation of these processes is required before questions pertaining 
to evolutionary connections among dwarf irregulars (dIs), dwarf ellipticals (dEs) and
BCDs (Thuan \cite{thuan85}, Davies \& Phillipps \cite{davies88},  Papaderos et al. 1996b; 
hereafter \cite{papaderos96b}, Patterson \& Thuan \cite{patterson96}, Salzer \& Norton 
\cite{salzer98}, Marlowe et al. \cite{marlowe97}, \cite{marlowe99}) can be answered.
Deep imaging of the low surface brightness (LSB) component of BCDs, 
discovered first by Loose \& Thuan (1985; hereafter \cite{lt85}), disclosed an 
evolved stellar population underlying the regions of active star formation (SF)
(Loose \& Thuan \cite{lt86}, Kunth et al. \cite{kunth88}, Papaderos et al. 1996a; 
hereafter \cite{papaderos96a}). 
Therefore, the initial hypothesis that BCDs are ``extragalactic \ion{H}{ii} regions''
lacking an older stellar population (Sargent \& Searle
\cite{sargent70}) had to be dismissed for the majority of these systems.

An exception to this finding is made by a tiny fraction ($\la$ 1\% ) of young galaxy  
candidates which recently have been identified among BCDs (Thuan et al. \cite{til97}, 
Papaderos et al. \cite{papaderos98c}, Thuan \& Izotov \cite{ti99}). 
There is increasing observational evidence that such systems, invariably found 
among the most metal--deficient ($Z\la$ 1/20 $Z_{\odot}$, Thuan \& Izotov 
\cite{ti99}, Izotov \& Thuan \cite{it99}) 
members of the BCD class, have started forming the bulk
of their stellar component less than 100 Myr ago.
This qualifies them as local counterparts of primeval galaxies, 
i.e. the first galactic building blocks which are thought to have 
formed at high redshifts.

It seems meaningful to 
search among the variety of dwarf galaxies for possible successors of young galaxies, 
i.e. objects in a more evolved stage with a substantial fraction of their masses 
still in gaseous form. An age sequence built therefrom would allow to study the 
evolution of nearby dwarf galaxies, which may at the same time give clues to 
early galaxy evolution processes at high redshift.
This is of great interest for the assessment of the starburst--driven 
evolution of young galaxies, some of which may become the building blocks
of larger systems and for the understanding of issues like the 
faint blue galaxy excess at intermediate redshifts 
(Babul \& Rees \cite{babul92}, Koo et al. \cite{koo97}, 
Guzm\'an et al. \cite{guzman98}).

Loose \& Thuan (1985) developed a morphological classification scheme for BCDs and 
found that $\sim$ 90\%  of their sample is made up of iE-- and nE--objects. These 
show, respectively, an irregular or nuclear starburst component superposed  on a 
smooth elliptical LSB host galaxy with optical colours pointing to ages of several 
Gyr (Loose \& Thuan \cite{lt86}, \cite{papaderos96b}, Telles \& Terlevich 
\cite{telles97}, Noeske \cite{noeske99}). 
In contrast, the few members of the rare class of i0 BCDs examined so far 
have been found to be young galaxy candidates (Thuan et al. 1997, Papaderos et al. 
\cite{papaderos98c}, Thuan \& Izotov \cite{ti99}, Izotov et al. \cite{izotov99}), 
as they apparently lack a smooth stellar underlying component that would have 
required previous episodes of SF to be built.

\begin{figure*}[t]
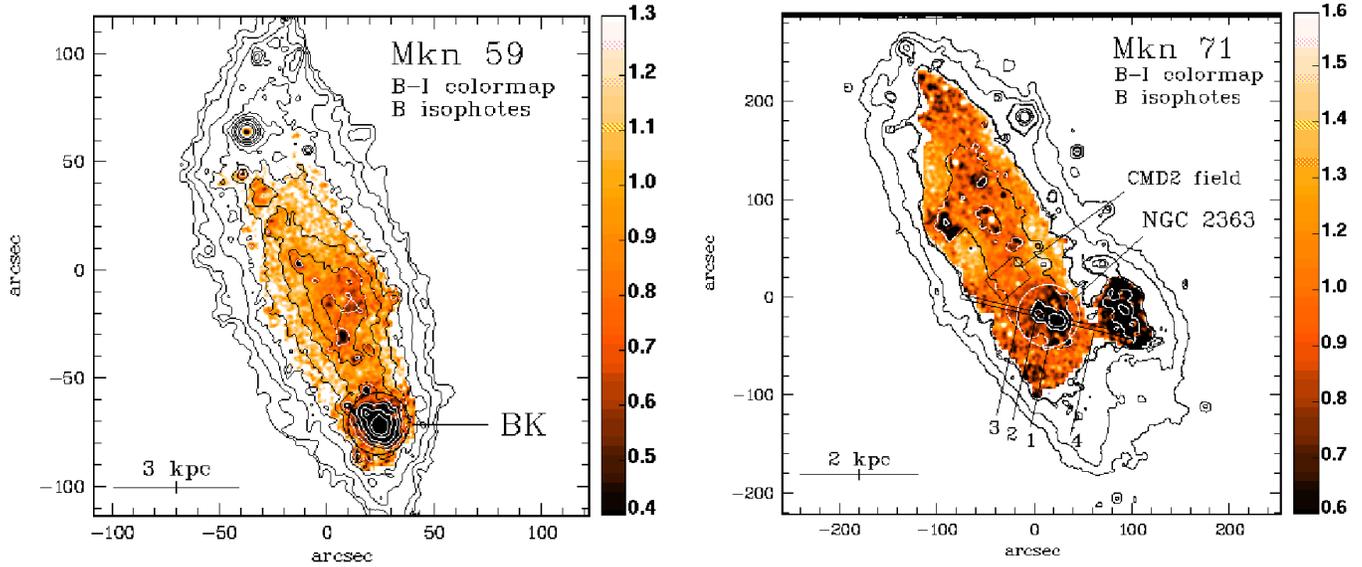

\centerline{
\psfig{figure=9237.f1,height=7.5cm,clip=}\hspace*{0.5cm}
\raisebox{1.5mm}{\psfig{figure=9237.f2,height=7.32cm,clip=}}
}
\caption{$(B-I)$ colour maps with overplotted $B$ band isophotes. 
{\bf left} \object{Mkn\,59}; the isophotes correspond to surface brightness levels of 20, 21, 22 and 
22.5--25.5 mag/$\sq\arcsec$ in steps of 0.5 mag. The bright starburst knot (BK), described in detail 
by Dottori et al. (\cite{dottori94}) is indicated. 
{\bf right} \object{Mkn\,71}; the isophotes correspond to surface brightness levels of 
21, 22, 23, 24, 24.5 and 25 mag/$\sq\arcsec$. The giant \ion{H}{ii} complex 
\object{NGC\,2363}, as well as the field used to derive the colour--magnitude 
diagram of the galaxy's underlying stellar population from HST data (CMD\,2) are marked.
The indices 1--4 along the orientation of the long--slit (centered at the axis origin) mark the 
regions from which the spectra displayed in Figure\ 8 were extracted. 
North is up and east to the left.}
\label{bi_maps}
\end{figure*}

Only the class of iI BCDs, characterized by both irregular LSB components and 
starburst regions, has not yet been studied systematically.
This paper focuses on the subset of iI,C BCDs (\cite{lt85}) where ``C'' denotes a 
\emph{cometary} appearance caused by a bright star--forming complex at one end 
of an elongated stellar galactic body (see Figs. \ref{bi_maps} and \ref{mk59_bslit}).
%
%---------------------------- Figure1 ---------------------------------
%------------------------------------------------------------------
%
In the notation of Salzer et al. (\cite{salzer89}) an iI,C BCD is to be included in 
the class of \ion{H}{ii} hotspot galaxies; these systems seem to systematically 
differ from other subclasses of \ion{H}{ii} galaxies with respect to the 
metallicities and excitation properties of their ionized gas. 
\begin{table}
\caption[]{\label{obs_ca} Ground--based imaging}
\begin{tabular}{lcrcll} 
\hline
Object 		&Filter	& $t_{\rm exp}$& Night & Seeing 		& Sky 	\\
$\alpha$(2000.0)&       & sec      &       & \arcsec\,fwhm	& mag$/\sq\arcsec$ \\ 
$\delta$(2000.0)& & & & & \\
(1) & (2) & (3) & (4) & (5) & (6) \\ \hline
\object{Mkn\,59} 	& $B$         & 180  & 4	&  1.67  & 22.4 \\
          		& $B$         & 1100 & 3	&  2.95  & 22.5 \\
12 59 00.3		& $R$         & 540  & 4  &  1.57  & 20.8 \\
+34 50 43 		& $I$         & 300  & 4	&  1.45  & 18.8 \\ \hline
\object{Mkn\,71} 	& $B$         & 1500 & 1	&  1.59  & 22.1 \\ 
          		& $R$         & 300  & 4	&  1.48  & 20.6 \\ 
7 28 41.4 		& $R$         & 600  & 3 	&  2.92  & 20.7 \\ 
+69 11 26 		& $I$         & 600  & 1	&  1.67  & 18.9 \\ 
          		& H$\alpha$ & 900  & 3	&  3.07  & ---  \\ 
          		& H$\alpha$ & 900  & 4	&  1.74  & ---  \\ \hline
\end{tabular}%
\end{table}
While the main bodies of iI,C systems are reminiscent of low surface brightness 
dwarf irregulars (dIs) or magellanic irregulars (Kennicutt et al. \cite{kennicutt80}, 
Dottori et al. \cite{dottori94}, Wilcots et al. \cite{wilcots96}) with some 
widespread low--level SF, the bright off--center \ion{H}{ii} regions 
are typical for BCDs with respect to their sizes, H$\alpha$ luminosities and 
electron temperatures. The detection of Super Star Clusters (SSCs) in these
spots points to a very recent or still ongoing starburst episode 
(Kennicutt et al. \cite{kennicutt80},  
Dottori et al. \cite{dottori94}, Barth et al. \cite{barth94}).

Optical surveys of BCDs (\cite{lt85}, Kunth et al. \cite{kunth88}, Salzer et
al. \cite{salzer89}, \cite{papaderos96a}, Marlowe et al. \cite{marlowe97}, Doublier et al. 
\cite{doublier97}) and dIs (e.g. Hunter et al. \cite{hunter98}) suggest that
in these systems SF per unit area correlates with  
the local surface density of the underlying host galaxy.
In this respect, the occurrence of a starburst with the amplitude
observed in iI,C BCDs at one end of an elongated stellar LSB body 
appears puzzling and may be interpreted in at least two ways:

(i) iI,C BCDs are in fact dIs or magellanic irregulars caught during a brief 
stochastic enhancement of their otherwise moderate star formation rate per 
unit area. This would imply that both the structural properties of their stellar 
hosts and \ion{H}{I} envelopes will be indistinguishable from those of other dIs.

(ii) In iI,C BCDs the conditions necessary for the ignition of a starburst are 
not fulfilled at the center but in the outskirts of their LSB components. 
This may be due to a conspiracy of intrinsic and external properties of an 
evolved BCD (kinematics of the gaseous component and Dark Matter distribution 
or external perturbation by a companion or intracluster gas) or alternatively 
be a common feature among less evolved systems.

Here we attempt to test such hypotheses by analysing optical data of 
two nearby iI,C BCDs, \object{Mkn\,59} (\object{NGC\,4861}) and \object{Mkn\,71}
(\object{NGC\,2366}, hosting the \ion{H}{ii} region complex \object{NGC\,2363}).
The distance to \object{Mkn\,71} was determined to $D$=3.44\,Mpc through 
observations of Cepheids (Tolstoy et al. \cite{tolstoy95}). 
For \object{Mkn\,59} different models for infall and small--scale 
perturbations in the Virgo Cluster environment yield a distance  
between $\sim $ 10.7 Mpc (Heckman et al. \cite{heckman98}) and 17.8 Mpc
(Tully \cite{tully88}). 
The principal results of this work do not depend on the exact distance 
to \object{Mkn\,59} (cf. Section \ref{res_struct}). 
Throughout this paper, we adopt $D$=10.7\,Mpc (Heckman et al. \cite{heckman98}). 

Using broad-- and narrow band images and long--slit spectra from own ground--based observations 
and archival HST data, we shall investigate the physical state and the chemical composition 
of the ISM and examine whether the iI,C BCDs under study differ from typical
iE/nE BCDs with respect to the ages and structural properties of their LSB hosts.
In Section\ 2, we describe our observations and data analysis. 
We present our results in Section 3, and discuss them in Section 4.
In Section 5, we summarize our results and conclusions.
%
% =======================================
\section{Data acquisition and processing}
% =======================================
%
\subsection{Ground--based imaging}
\subsubsection{Observations and data reduction}
\label{obred_ca}
%
% ---------- Figure2--------------------------------------
%
\begin{figure*}[t]
\resizebox{12cm}{!}{\psfig{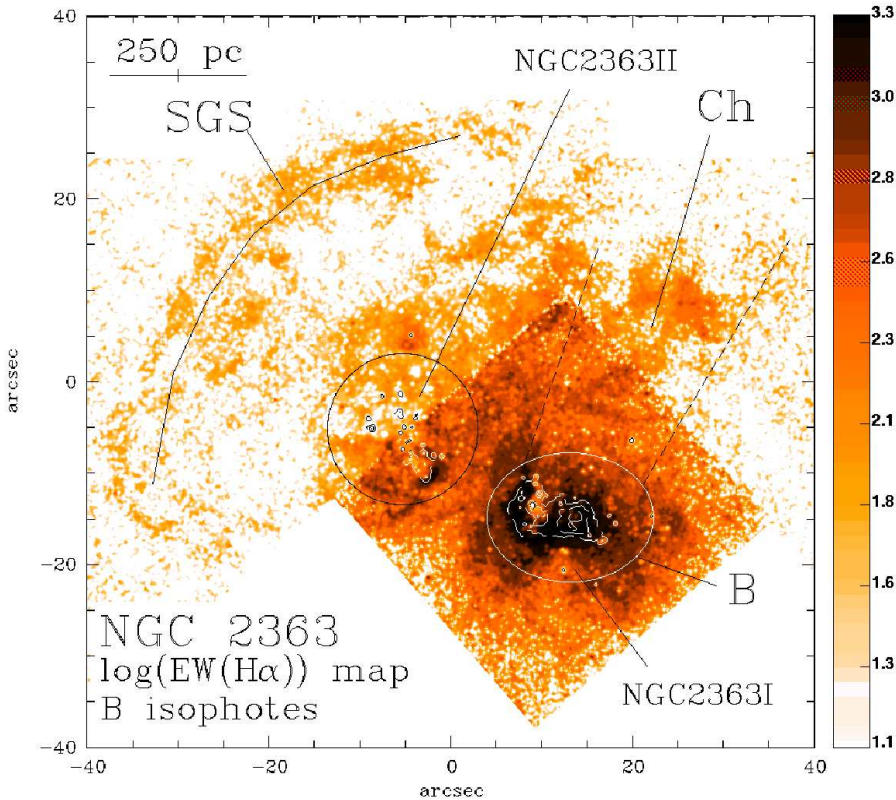}}
\hfill
\parbox[b]{55mm}{
\caption{
Logarithmic representation of the H$\alpha$ equivalent width
$\left(EW({\rm H}\alpha){\rm [\AA]}\right)$ map of the \ion{H}{ii} complex 
\object{NGC\,2363} within \object{Mkn\,71},
computed from HST/WFPC2 images as described in Section\ 2.1.1. 
$B$ band contours at 18--20 mag/$\sq\arcsec$ mark the positions 
of the young star clusters \object{NGC\,2363\,I} and II within \object{NGC\,2363}.
The arc--like structure, extending out to $\sim$ 0.5\,kpc northeast of 
\object{NGC\,2363\,II}, probably marks a supergiant shell
(SGS) (Hunter \& Gallagher \cite{hunter97}). 
 The expanding superbubble (B) and the probable outflow chimney (Ch, delimited by 
the pair of dashed lines) around \object{NGC\,2363\,I} (Roy et al. \cite{roy91}) 
are indicated.
The spatial distribution of the young stellar background in \object{NGC\,2363\,I} and 
\object{NGC\,2363\,II} is reviewed in Section \ref{metal_discuss}.
North is up and east to the left.
\label{mk71_hb} } }
\end{figure*}
%---------------------------------------------------------------

Images were taken on March 7th -- 10th 1997 at the 2.2m telescope of the 
German--Spanish Astronomical Center, Calar Alto, Spain. We used the Calar Alto Faint Object 
Spectrograph ({CAFOS}), equipped with a 2048$\times$2048 pixel SITe\#1d CCD.
The focal ratio of f/4.4 in the RC focus and the pixel size of 24 $\mu$m/pixel yield 
an instrumental scale of 0.53\arcsec/pixel and a usable field of view of $\sim$  15\arcmin. 
At a gain ratio of 2.3\,e$^-$\,ADU$^{-1}$ the read--out noise was $<$ 3 counts (rms).
Column\ (1) of the observing log (Table \ref{obs_ca}) contains the 
equatorial coordinates of the targets, cols.\ (2) and (3) give 
the filters and exposure times, respectively. 
Column\ (4) lists the night of the observing run, starting with night 1 from 
March 7th to March 8th 1997. The seeing and the mean sky surface brightness 
during each exposure are given in cols. (5) and (6).
During each night, dark--, bias--, and flat--field exposures were taken, and the photometric 
standard field \object{NGC\,2419} (Christian et al. \cite{christian85}) was observed at different 
zenith angles. 
Using the ESO MIDAS\footnote{Munich Image Data Analysis System, provided by the 
European Southern Observatory (ESO).} software package, standard reduction and calibration steps 
were applied to the raw images. During nights 3 and 4 the conditions were photometric, resulting 
in calibration errors well below 0.05\,mag. A poor atmospheric transparency during night 1 led to 
strong airmass--dependent terms for the $B$ band exposure of \object{Mkn\,71}. 
Therefore the calibration was accomplished by comparing aperture photometry of bright isolated 
point sources on HST-- and ground--based images. The photometric uncertainty 
of the latter $B$ band exposure was found to be $\la$ 0.1\,mag. 
All exposures for one object were aligned to each other using positions of 
point sources; the transformed images were generated by re--sampling the
original image using a flux--conserving routine.
Colour maps were derived after the resolution of different frames had been 
equalized by convolving the exposure with the better resolution with a 
normalized gaussian distribution of adequate width.

The continuum level in the H$\alpha$ images was inferred by scaling the $R$ band images by an 
empirically determined factor $C$ so that the fluxes of field stars were matched between the raw 
H$\alpha$-- and the scaled $R$ exposures. In turn, the scaled $R$ band exposures were  
subtracted from the raw H$\alpha$ images, giving the emission flux $F_{\rm em}$ at each pixel. 
The latter approach is strictly correct only when the H$\alpha$ emission line is contributing a 
minor fraction of the line--of--sight $R$ band flux. In regions where H$\alpha$ emission makes 
a substantial fraction of the photons received in the $R$ band the empirical scaling may lead to 
a slight overestimation of the continuum flux. 
In these cases the correct H$\alpha$ emission $F_{em}^{c}$ is obtained from 
$F_{em}$ as:
\begin{equation}
\label{hacorr}
F_{em}^{c}=\frac{1}{1-C\cdot T_{R}({\rm H}\alpha )}F_{em},  
\end{equation} 
where $C$ is the empirical factor described above and $T_{R}({\rm H}\alpha )$ 
the mean transmission of the $R$ band filter at the H$\alpha$ wavelength.
The resulting narrow band frames were calibrated using continuum--subtracted H$\alpha$ exposures 
of the planetary nebula \object{NGC\,2392} for which aperture flux measurements are given by 
Kaler (\cite{kaler83}).

To derive H$\alpha$ equivalent width maps we subtracted first the corrected line emission 
frame $F_{em}^{c}$ (Eq.\ref{hacorr}) from the raw H$\alpha$ image to obtain a continuum image 
$F_{cont}$, which was then normalized to the continuum flux per 1\AA\  wavelength interval by 
dividing it by the effective width of the H$\alpha$ filter.
The emission line image $F^{c}_{em}$ was divided by the transmission of the H$\alpha$ filter 
at the H$\alpha$ wavelength.
The resulting frame, containing the total H$\alpha$ emission flux, was divided by the 
1\AA -- normalized continuum flux frame to obtain the H$\alpha$ equivalent width image.

The colour- and H$\alpha$ maps were corrected for interstellar extinction 
following Savage \& Mathis (\cite{savage79}) and using the extinction
coefficients $C$(${\rm H}\beta$) given in Section \ref{spectr_obs}, which
translate into $E(\!B\!-\!V\!)$=0.06 mag and $E(\!B\!-\!V\!)$=0.08 mag for 
\object{Mkn\,59} and \object{Mkn\,71}, respectively.
%
%------------------------------------------
\subsubsection{Surface photometry}
%------------------------------------------
%
% ---------- Figure3 -----------------------------------------
\begin{figure*}[t]
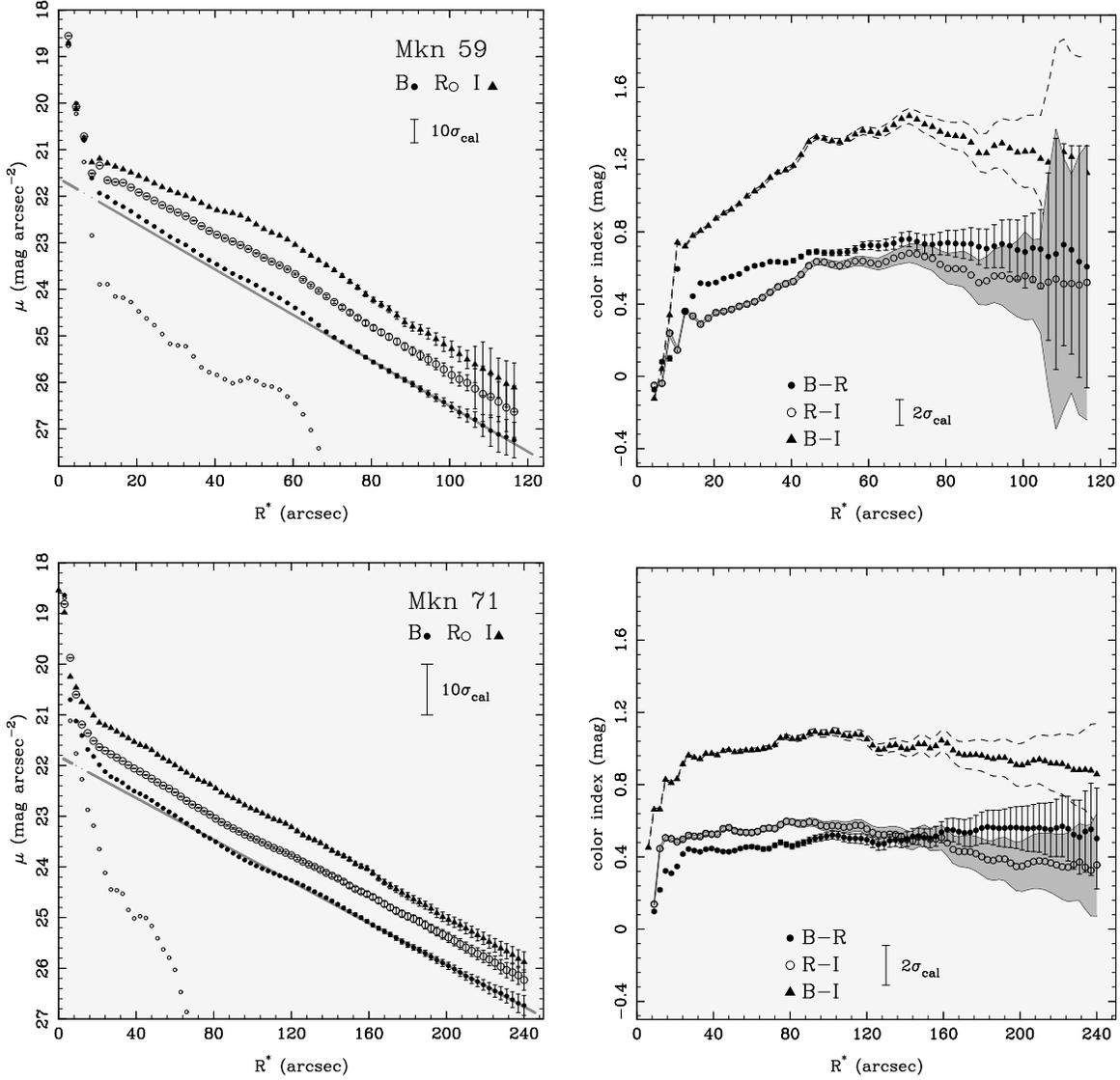

\centerline{\psfig{figure=9237.f4,height=7.2cm,angle=-90,clip=}\hspace*{0.6cm}\psfig{figure=9237.f5,height=7cm,angle=-90,clip=}}\ \\
\centerline{\psfig{figure=9237.f6,height=7.2cm,angle=-90,clip=}\hspace*{0.6cm}\psfig{figure=9237.f7,height=7cm,angle=-90,clip=}}
\caption{
{\bf left} Surface brightness profiles of \object{Mkn\,59} (top) and \object{Mkn\,71} (bottom) in $B$, $R$ and $I$. 
Small circles show the surface brightness distribution of the residual $B$ band emission in 
excess of the exponential fit to the LSB component (straight line). 
The calibration uncertainty $\sigma _{cal}$ effecting a vertical shift of the entire profile is displayed
10 times enlarged. 
{\bf right} Radially averaged colour profiles with 2 times enlarged calibration error bars.
Note that for surface brightness levels fainter than $\sim 23.5$ $B$ mag arcsec$^{-2}$ the 
colour index of the underlying LSB galaxy becomes nearly constant.} 
\label{profs}
\end{figure*}
%-------------------------------------------------------------------
%
\begin{table*}[ht]
\caption{\label{decomp_res}Structural properties of the starburst-- and LSB components of \object{Mkn\,59} and \object{Mkn\,71}
derived from profile decomposition.}
\begin{tabular}{lccccccccc}
\hline
Name & Band & $\mu _{E,0}$ & $\alpha $ & $P_{25}$  & $m_{P_{25}}$ & $E_{25}$ & $m_{\rm E_{25}}$ & $m_{tot}$
 & $r_{eff}$,$r_{80}$\\
     &      & mag/$\sq\arcsec$ & kpc       & kpc       &  mag             &  kpc     &  mag             & mag      &  kpc  \\
 (1) &  (2) &   (3)            &   (4)     &  (5)      &   (6)            &   (7)    &  (8)             &  (9)     &  (10) \\
\hline
\object{Mkn\,59}  & $B$ & 21.62$\pm$0.15  & 1.151$\pm$0.041  & 1.39  & 13.96  & 3.59  & 13.11  & 12.64$\pm$0.01  & 1.37, 2.83\\
       		  & $R$ & 20.78$\pm$0.23  & 1.123$\pm$0.059  & 0.37  & 14.11  & 4.36  & 12.22  & 12.08$\pm$0.02  & 1.67, 3.07\\
       		  & $I$ & 19.77$\pm$0.24  & 1.022$\pm$0.051  & 0.17  & 14.02  & 4.92  & 11.36  & 11.64$\pm$0.02  & 2.01, 3.29\\
\hline
\object{Mkn\,71}  & $B$ & 21.82$\pm$0.12  & 0.879$\pm$0.028  & 0.65  & 13.39  & 2.58  & 11.47  & 11.12$\pm$0.02  & 1.19, 2.25 \\
         	  & $R$ & 21.29$\pm$0.15  & 0.885$\pm$0.035  & 0.50  & 13.51  & 3.02  & 10.84  & 10.68$\pm$0.03  & 1.28, 2.34 \\
         	  & $I$ & 20.57$\pm$0.15  & 0.816$\pm$0.030  & 0.29  & 14.26  & 3.33  & 10.22  & 10.20$\pm$0.03  & 1.32, 2.30 \\
\hline
\end{tabular}
\end{table*}

Surface brightness profiles (SBPs) are the result of a transformation of a galaxy's 
2--dimensional flux pattern into a monotonically
decreasing 1--dimensional intensity distribution.
For a set of methods to compute the photometric radius $R^{\star}(\mu)$, i.e. the radius of a 
circle with an area equal to the one enclosed by the isophote at the surface brightness level 
$\mu [{\rm mag}/\sq\arcsec]$, see e.g. Loose \& Thuan (1986) and \cite{papaderos96a}.
The method of summing up the area defined by pixels with fluxes exceeding a threshold $I$($\mu$) 
(method {\sf iii}; \cite{papaderos96a})
was considered most appropriate for the BCDs analyzed here as both their underlying and starburst 
components are of irregular morphology. Other techniques, such as isophote integration or ellipse 
fitting, though well suited for more regular BCDs, are hardly applicable to the optical images of 
\object{Mkn\,59} and \object{Mkn\,71}, mainly due to the
fact that their high surface brightness components split into many star-forming regions 
occupying a substantial fraction of the host galaxy's surface.

For each surface brightness level $\mu$, the number $N(\mu)$ of all pixels within a polygonal 
aperture with a count level $\ge I(\mu )$ was determined; the corresponding photometric radius is
\begin{equation}
R^{\star}(\mu)=\frac{1}{\pi}\sqrt{N(\mu)\cdot A_{\rm pxl}},
\end{equation} 
where $A_{\rm pxl}$ is the solid angle per pixel in $\sq\arcsec$.
Numerical simulations showed that for the technique described above
Poisson photon noise may cause a systematic flattening of SBPs 
at a signal to noise $S/N$ level $\la 10$. 
This frequently observed artificial profile flattening could be, however, 
satisfactorily corrected down to a $({S/N})$ level of $\approx 1$ 
by using the adaptive filtering algorithm implemented in MIDAS
(see Richter et al. 1991 for a description).
The uncertainties at each photometric radius were calculated taking into account the corresponding 
intensity, the sky noise of each image, and the number of pixels involved in its calculation 
(cf. \cite{papaderos96a}). 
Colour profiles were computed by subtracting SBPs from each other; 
the surface brightness-- and colour profiles displayed in Figure\ 3 are corrected for 
interstellar extinction (cf. Section\ \ref{obred_ca}).\smallskip    

Unlike stellar systems like globular clusters and giant ellipticals where the mass to light 
($M/L$) ratio may be regarded nearly constant over the whole system, the integrated luminosity 
of BCDs originates from two distinct stellar populations with substantially different ages and 
$M/L$ ratios.
Although the LSB component contributes on average one half of the total $B$ band luminosity
of a BCD (\cite{papaderos96b}, Salzer \& Norton 1998), it contains most of the stellar mass of 
the system. 
Therefore, its intensity distribution yields information on the inner gravitational 
potential within which the starburst ignites and evolves. 
In order to disentangle the light distribution of the older host galaxy from the one of the 
superimposed starburst component, we applied a simple 2--component decomposition scheme adjusted 
interactively to each profile, instead of iteratively fitting the full nonlinear 3--component 
scheme described in \cite{papaderos96a}. The intensity distribution of the LSB component is 
approximated by an exponential fitting law of the form
\begin{equation}
I_{\rm E}(R^{\star})=I_{\rm E,0}\exp\left(-\frac{R^{\star}}{\alpha}\right),
\end{equation}
or equivalently 
\begin{equation}
\mu (R^{\star}) = \mu _{\rm E,0} + 1.086 \left(\frac{R^{\star}}{\alpha}\right)
\label{linfit}
\end{equation}
where $\alpha$ denotes its exponential scale length in arcsec and $\mu$ the
surface brightness level in mag/$\sq\arcsec$. 
Equation (\ref{linfit}) was adjusted to each profile 
by applying an error weighted linear fit to the data,
at radii sufficiently large to be free from the contamination of the starburst light, i.e. 
where the colour profiles become constant and H$\alpha$ emission vanishes. 
To check at which surface brightness levels gaseous emission becomes 
negligible, we overlaid the H$\alpha$ emission-- and $EW($H$\alpha)$ maps with broad band 
isophotes (cf. Figs. \ref{bi_maps} and \ref{mk71_hb}) .
The fits, obtained at radii $R^{\star} \geq$\,76$\arcsec$ for \object{Mkn\,59} and 
$R^{\star} \geq$\,160$\arcsec$ for \object{Mkn\,71}, yield the extrapolated central 
surface brightness $\mu _{E,0}$ and the exponential scale length $\alpha$. 
The radial surface brightness distribution of the starburst 
component can be computed from the residual luminosity in excess of the fit Eq.\ (\ref{linfit}).

Table \ref{decomp_res} summarizes the results of the profile decomposition. 
In the same way as described in \cite{papaderos96a}, cols. (3) and (4) give the central 
surface brightness and exponential scale length of the LSB component. 
Columns (5) and (7) list respectively the isophotal radii of the starburst component
$P_{25}$ (practically the ``plateau radius'' used in \cite{papaderos96a}), 
and of the stellar LSB host, $E_{25}$. Both radii were determined from
extinction--corrected SBPs at a surface brightness level of 25 mag/$\sq\arcsec$. 
Columns (6) and (8) contain the apparent magnitudes of the latter components determined within 
$P_{25}$ and $E_{25}$, respectively, and col. (9) the total apparent magnitude in each band 
as derived from integration of the corresponding SBP out to the last measured point. 
Column (10) lists the effective radius $r_{eff}$ and the radius $r_{80}$, enclosing 80\% 
of the galaxy's total flux.
%
%--------------------------------------------------------------
\subsection{HST images}
%--------------------------------------------------------------
%
% ---------- Figure4-------------------------------------------
\begin{figure*}[t]
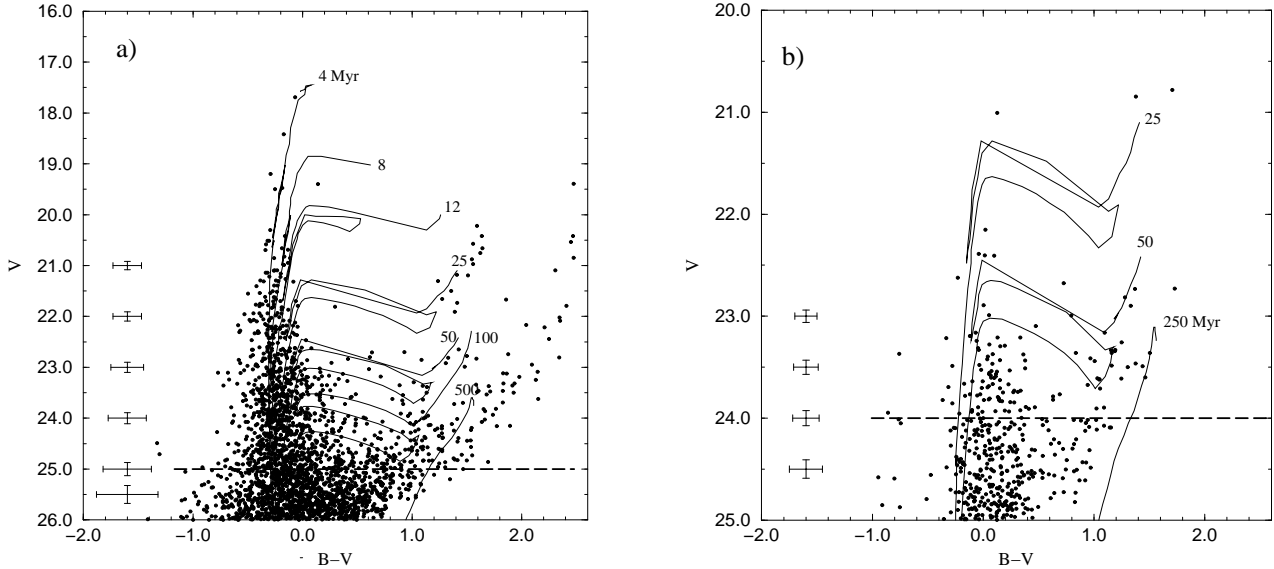

\centerline{\psfig{figure=9237.f8,width=9cm}\psfig{figure=9237.f9,width=9.1cm}}
\caption{$V$ vs. $(B-V)$ colour--magnitude diagrams (CMDs) of \object{Mkn\,71}; overplotted are isochrones by Bertelli et al. (\cite{bertelli94}) for a metallicity of 1/20\ $Z_{\odot}$. 
Crosses to the left illustrate the mean errors in both axes as a function of the apparent $V$ magnitude; the dashed lines indicate the estimated confidence limit for the $B$ and $V$ band. 
{\bf a} CMD derived within the PC1 chip of WFPC2, centered on the \ion{H}{ii} region \object{NGC\,2363};
the isochrones correspond to ages of 4, 8, 12, 25, 50, 100 and 500 Myr. 
{\bf b} CMD of the region of the main body of \object{Mkn\,71} indicated in Figure\ 1 (right panel, see label ''CMD 2'')
where active star--forming regions are absent as derived from WF4 data.
The isochrones correspond to ages of 25, 50 and 250 Myr.}
\label{cmds}
\end{figure*}
%--------------------------------------------------------------
\subsubsection{Data reduction}
%--------------------------------------------------------------
Next we shall discuss colour--magnitude diagrams (CMDs) of the \ion{H}{ii} complex 
\object{NGC\,2363} and the main body of \object{Mkn\,71} derived from Hubble Space Telescope 
(HST) archival data.
The following analysis is based on integrations of 2$\times$800 and 1$\times$700 sec in the 
F439W filter, 2$\times$800 sec in the F547M filter and 1$\times$900 plus 1$\times$600 sec in the 
F656N filter taken on January 8th, 1996 with the WFPC2 (PI: Drissen, Proposal ID 06096). 
All images were reduced through the standard pipeline as described in Holtzman 
et al. (\cite{holtzman95a}). Exposures taken in the same filter were co--added and corrected 
for cosmic ray events using the STSDAS package and IRAF\footnote[2]{IRAF is distributed by the
National Optical Astronomy Observatories, which is operated by the
Association of Universities for Research in Astronomy, Inc., under
cooperative agreement with the National Science Foundation.}. 
We applied a charge transfer efficiency 
correction to the data as described in the HST Data Handbook (\cite{hst98}), performed a sky 
background subtraction and used the synthetic zero points and transformation coefficients given 
by Holtzman et al. (\cite{holtzman95b}) to transform the measured fluxes to magnitudes in the 
Johnson $UBVRI$ system.
For the H$\alpha$ narrow band filter, F656N, the continuum was subtracted as described in 
Section \ref{obred_ca}; the correction given by Eq. (\ref{hacorr}) was not necessary, as the 
H$\alpha$ line lies outside the wavelength range covered by the F547M filter. The calibration of 
the H$\alpha$ images was done following the prescriptions by Holtzman et al. (\cite{holtzman95b}).
%--------------------------------------------------------------
%
\subsubsection{Colour--magnitude diagrams}
\begin{figure}[t]
\resizebox{\hsize}{!}{\includegraphics{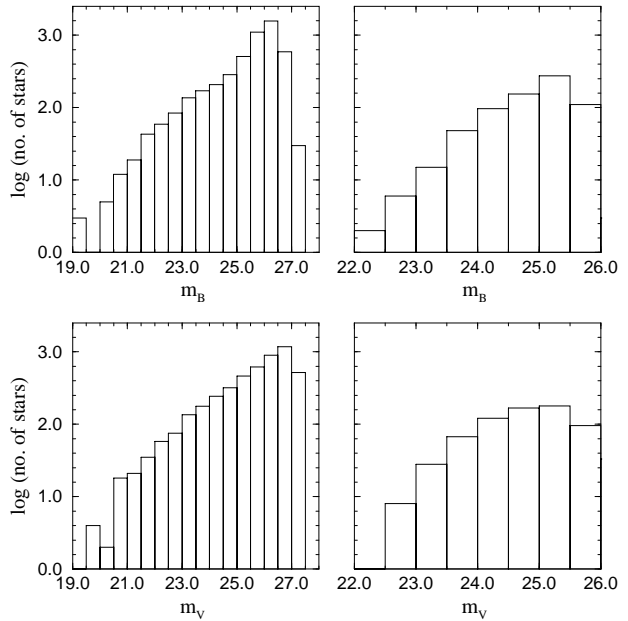}}
\caption{Luminosity functions of the point sources included in the colour--magnitude diagrams (Figure \ref{cmds}) as
obtained within the PC1-- (left) and WF4 chip (right) of HST WFPC2.}
\label{compl_pc}
\end{figure}

The \ion{H}{ii} region \object{NGC\,2363} within \object{Mkn\,71} (Figure\,1; right) was centered 
on the PC/WF1 chip
(instrumental scale of 0\farcs 046/{pixel}), allowing the resolution of  compact groups of stars 
with a mean linear separation of $\sim$ 0.8 pc. 
We utilized the PC1 data to derive CMDs in order to further constrain the results of the spectral
population synthesis analysis (Section 3.3) using the DAOPHOT\,II stellar photometry 
package under ESO MIDAS. 
A model for the point spread function (PSF), necessary for performing multiple--PSF fitting in 
crowded fields, was computed from several isolated bright stars in each frame. 
Because of the severe crowding within some regions of \object{NGC\,2363} as well as local 
background variations due to strong gaseous emission, we evaluated the photometric uncertainties 
and completeness limits  of the resulting point-source identification files, 
taking into account the error growth towards fainter luminosities and the magnitude histograms 
(cf. Figure \ref{compl_pc}).

The measured fluxes were calibrated and transformed to the Johnson $B$ and $V$ filters as described 
in the previous section. 
The luminosity functions of the detected point sources (Figure \ref{compl_pc}, left panel) suggest 
a reasonable completeness for sources brighter than $\approx$ 26 mag in both $B$ and $V$. 
Faintwards of the latter magnitude cutoff the photometric errors become dominant (see Figure 
\ref{cmds}). 
Close to the detection limit, an additional error source is introduced by the small--scale 
variations of the background due to gaseous emission as well as strong noise peaks which possibly 
are misinterpreted as faint stars (cf. the excess counts at low luminosities in the upper left 
panel of Figure \ref{compl_pc}).
We shall therefore consider the data reliable down to a limit of $\sim$ 25 mag in $B$ and $V$,
close to the one Drissen et al. (\cite{drissen99}) obtain for the same data set.
This cutoff is sufficient for our purposes; from a simple comparison of data points and isochrones 
it becomes evident that, given the photometric errors and poor time resolution of the $(B-V)$ CMD, 
no firm conclusions on the SF history can be drawn for ages $\ga$ 50 Myr.
A second CMD was derived in the same way for a region on the WF4 chip apparently free of any 
appreciable signatures of current SF (marked in Figure \ref{bi_maps}) to obtain information on the 
galaxy's underlying stellar population separately. From an inspection of the right panels of 
Figure \ref{compl_pc}, and from the considerations referring to the CMD of \object{NGC\,2363} we 
estimate the confidence limit to $\approx $24 mag for both $B$ and $V$. 
Either CMD is shown in Figure\ \ref{cmds} with typical photometric uncertainties for different 
magnitude intervals along with synthetic isochrones for a metallicity of $Z$=0.001 adopted from 
Bertelli et al. (\cite{bertelli94}). 
%
%
% ---------- Figure5 ---------------------------------------
%
% ======================================================
\subsection {Spectroscopic observations and data reduction}
% ======================================================
\label{spectr_obs}
%--------------------------------------------------------------
%
% Table 1 (line intensities)
%
\begin{table*}[t]
\caption{Emission Line Intensities in \object{Mkn\,59}}
\label{Tab1}
\begin{tabular}{lccccc} \\ \hline 
Ion & \multicolumn{2}{c}{region 1}&        &
\multicolumn{2}{c}{region 2} \\ \cline{2-3} \cline{5-6}
    &$F$($\lambda$)/$F$(H$\beta$)
&$I$($\lambda$)/$I$(H$\beta$)&     
&$F$($\lambda$)/$F$(H$\beta$)
&$I$($\lambda$)/$I$(H$\beta$) \\ \hline 
 3727\ [\ion{O}{ii}]        &1.0086$\pm$0.0018& 1.0879$\pm$0.0020&&1.9191$\pm$0.0832&2.0801$\pm$0.1049 \\ 
 3750\ H12 	            &0.0204$\pm$0.0006& 0.0268$\pm$0.0009&&  ...            &  ...             \\ 
 3770\ H11        	    &0.0280$\pm$0.0006& 0.0349$\pm$0.0009&&  ...            &  ...             \\ 
 3798\ H10        	    &0.0341$\pm$0.0006& 0.0415$\pm$0.0009&&  ...            &  ...             \\ 
 3835\ H9            	    &0.0498$\pm$0.0006& 0.0581$\pm$0.0009&&  ...            &  ...             \\ 
 3868\ [\ion{Ne}{iii}]      &0.4566$\pm$0.0011& 0.4872$\pm$0.0012&&0.3179$\pm$0.0503&0.3366$\pm$0.0584 \\ 
 3889\ \ion{He}{i} + H8     &0.1633$\pm$0.0008& 0.1786$\pm$0.0010&&  ...            &  ...             \\ 
 3968\ [\ion{Ne}{iii}] + H7 &0.2806$\pm$0.0009& 0.3014$\pm$0.0011&&0.1467$\pm$0.0309&0.2848$\pm$0.0698 \\ 
 4026\ \ion{He}{i}          &0.0110$\pm$0.0005& 0.0116$\pm$0.0005&&  ...            &  ...             \\ 
 4069\ [\ion{S}{ii}]        &0.0097$\pm$0.0006& 0.0102$\pm$0.0006&&  ...            &  ...             \\ 
 4076\ [\ion{S}{ii}]        &0.0031$\pm$0.0036& 0.0033$\pm$0.0038&&  ...            &  ...             \\  
 4101\ H$\delta$     &0.2396$\pm$0.0008& 0.2552$\pm$0.0010&&0.1296$\pm$0.0246&0.2606$\pm$0.0598 \\ 
 4340\ H$\gamma$     &0.4589$\pm$0.0011& 0.4766$\pm$0.0012&&0.3775$\pm$0.0316&0.4737$\pm$0.0461 \\ 
 4363\ [\ion{O}{iii}]       &0.0890$\pm$0.0006& 0.0916$\pm$0.0006&&0.0560$\pm$0.0326&0.0550$\pm$0.0349 \\ 
 4471\ \ion{He}{i}          &0.0357$\pm$0.0005& 0.0365$\pm$0.0005&&0.0875$\pm$0.0320&0.0846$\pm$0.0339 \\ 
 4658\ [\ion{}{iii}]      &0.0054$\pm$0.0004& 0.0054$\pm$0.0004&&  ...            &  ...             \\ 
 4686\ \ion{He}{ii}         &0.0131$\pm$0.0005& 0.0132$\pm$0.0005&&  ...            &  ...             \\ 
 4713\ [\ion{Ar}{iv}] + \ion{He}{i}&0.0084$\pm$0.0004& 0.0085$\pm$0.0004&&  ...            &  ...             \\ 
 4740\ [\ion{Ar}{iv}]       &0.0055$\pm$0.0004& 0.0056$\pm$0.0004&&  ...            &  ...             \\ 
 4861\ H$\beta$      &1.0000$\pm$0.0017& 1.0000$\pm$0.0017&&1.0000$\pm$0.0495&1.0000$\pm$0.0554 \\ 
 4921\ \ion{He}{i}          &0.0082$\pm$0.0004& 0.0081$\pm$0.0004&&  ...            &  ...             \\ 
 4959\ [\ion{O}{iii}]       &2.0670$\pm$0.0030& 2.0503$\pm$0.0030&&1.4993$\pm$0.0672&1.3588$\pm$0.0667 \\ 
 5007\ [\ion{O}{iii}]       &6.1585$\pm$0.0078& 6.0917$\pm$0.0078&&4.3242$\pm$0.1668&3.8956$\pm$0.1645 \\ 
 5199\ [\ion{N}{i}]         &0.0033$\pm$0.0004& 0.0033$\pm$0.0004&&  ...            &  ...             \\ 
 5271\ [\ion{Fe}{iii}]      &0.0019$\pm$0.0004& 0.0019$\pm$0.0003&&  ...            &  ...             \\ 
 5518\ [\ion{Cl}{iii}]      &0.0043$\pm$0.0004& 0.0041$\pm$0.0003&&  ...            &  ...             \\ 
 5538\ [\ion{Cl}{iii}]      &0.0026$\pm$0.0004& 0.0025$\pm$0.0004&&  ...            &  ...             \\ 
 5876\ \ion{He}{i}          &0.1138$\pm$0.0005& 0.1075$\pm$0.0005&&0.0854$\pm$0.0160&0.0698$\pm$0.0144 \\ 
 6300\ [\ion{O}{i}]         &0.0210$\pm$0.0004& 0.0195$\pm$0.0004&&  ...            &  ...             \\ 
 6312\ [\ion{S}{iii}]       &0.0209$\pm$0.0004& 0.0194$\pm$0.0003&&  ...            &  ...             \\ 
 6364\ [\ion{O}{i}]         &0.0072$\pm$0.0003& 0.0066$\pm$0.0003&&  ...            &  ...             \\ 
 6548\ [\ion{N}{ii}]        &0.0242$\pm$0.0005& 0.0222$\pm$0.0005&&  ...            &  ...             \\ 
 6563\ H$\alpha$     &3.0749$\pm$0.0041& 2.8214$\pm$0.0041&&3.6134$\pm$0.1384&2.8062$\pm$0.1277 \\ 
 6583\ [\ion{N}{ii}]        &0.0540$\pm$0.0004& 0.0495$\pm$0.0004&&0.1800$\pm$0.0229&0.1377$\pm$0.0194 \\ 
 6678\ \ion{He}{i}          &0.0323$\pm$0.0004& 0.0295$\pm$0.0003&&0.0427$\pm$0.0137&0.0324$\pm$0.0114 \\ 
 6717\ [\ion{S}{ii}]        &0.1052$\pm$0.0005& 0.0959$\pm$0.0005&&0.2983$\pm$0.0266&0.2258$\pm$0.0225 \\ 
 6731\ [\ion{S}{ii}]        &0.0785$\pm$0.0005& 0.0715$\pm$0.0004&&0.1597$\pm$0.0245&0.1207$\pm$0.0204 \\ 
 7065\ \ion{He}{i}          &0.0280$\pm$0.0003& 0.0252$\pm$0.0003&&  ...            &  ...             \\ 
 7136\ [\ion{Ar}{iii}]      &0.0862$\pm$0.0005& 0.0774$\pm$0.0004&&0.1331$\pm$0.0178&0.0974$\pm$0.0144 \\  
 7281\ \ion{He}{i}          &0.0051$\pm$0.0003& 0.0046$\pm$0.0002&&  ...            &  ...             \\ 
 7320\ [\ion{O}{ii}]        &0.0186$\pm$0.0004& 0.0166$\pm$0.0004&&  ...            &  ...             \\ 
 7330\ [\ion{O}{ii}]        &0.0209$\pm$0.0005& 0.0187$\pm$0.0005&&  ...            &  ...             \\  \\ 
 $C$(H$\beta$) dex    &\multicolumn {2}{c}{0.110$\pm$0.002}&&\multicolumn {2}{c}{0.235$\pm$0.050} \\ 
 $F$(H$\beta$)$^a$ &\multicolumn {2}{c}{15.08$\pm$0.02}&    &\multicolumn{2}{c}{0.28$\pm$0.01} \\ 
 $EW$(H$\beta$)\ \AA &\multicolumn {2}{c}{150.1$\pm$0.2}&&\multicolumn {2}{c}{44$\pm$1} \\ 
 $EW$(abs)\ \AA      &\multicolumn {2}{c}{0.35$\pm$0.04}&&\multicolumn{2}{c}{ 3.9$\pm$0.5} \\  \hline
\end{tabular}

$^a$in units of 10$^{-14}$ ergs\ s$^{-1}$cm$^{-2}$.
\end{table*}
                                                                                   
% Table 1a (line intensities)                                                           
%                                                                                           
%
\begin{table}
\caption{Emission Line Intensities in \object{Mkn\,59}}
\label{Tab1a}
\begin{tabular}{lccc} \\ \hline 
Ion & \multicolumn{2}{c}{region 3}& \\ \cline{2-3} 
    &$F$($\lambda$)/$F$(H$\beta$)
&$I$($\lambda$)/$I$(H$\beta$)&  \\ \hline 
 3727\ [\ion{O}{ii}]        &3.1730$\pm$0.1866& 2.9505$\pm$0.2018& \\                    
 3868\ [\ion{Ne}{iii}]      &0.5371$\pm$0.0516& 0.4994$\pm$0.0528& \\                    
 3968\ [\ion{Ne}{iii}] + H7 &0.2177$\pm$0.0343& 0.2832$\pm$0.0603& \\                    
 4101\ H$\delta$     	    &0.1568$\pm$0.0426& 0.2377$\pm$0.0836& \\                    
 4340\ H$\gamma$     	    &0.4335$\pm$0.0518& 0.4833$\pm$0.0714& \\                    
 4363\ [\ion{O}{iii}]       &0.0594$\pm$0.0287& 0.0552$\pm$0.0287& \\                    
 4861\ H$\beta$      	    &1.0000$\pm$0.0739& 1.0000$\pm$0.0848& \\                    
 4959\ [\ion{O}{iii}]       &1.3522$\pm$0.0939& 1.2574$\pm$0.0939& \\                    
 5007\ [\ion{O}{iii}]       &3.8327$\pm$0.2218& 3.5640$\pm$0.2219& \\                    
 6300\ [\ion{O}{i}]         &  ...            &   ...            & \\                    
 6563\ H$\alpha$    	    &2.5651$\pm$0.1519& 2.4329$\pm$0.1693& \\                    
 6583\ [\ion{N}{ii}]        &0.2177$\pm$0.0374& 0.2024$\pm$0.0378& \\                    
 6717\ [\ion{S}{ii}]        &0.3142$\pm$0.0410& 0.2921$\pm$0.0418& \\                    
 6731\ [\ion{S}{ii}]        &0.3037$\pm$0.0430& 0.2824$\pm$0.0437& \\                    
 7136\ [\ion{Ar}{iii}]      &  ...            &   ...            & \\  \\                   
 $C$(H$\beta$) dex    	&\multicolumn {2}{c}{0.000$\pm$0.077}	&\\        
 $F$(H$\beta$)$^a$ 	&\multicolumn {2}{c}{ 0.21$\pm$0.01}	&\\  
 $EW$(H$\beta$)\ \AA 	&\multicolumn {2}{c}{51.1$\pm$2.7}	&\\                  
 $EW$(abs)\ \AA      	&\multicolumn {2}{c}{3.85$\pm$1.49}	&\\ \hline
\end{tabular}
             
$^a$in units of 10$^{-14}$ ergs\ s$^{-1}$cm$^{-2}$.                                                    
\end{table}                                                                                    % 

\begin{figure}[t]
\centerline{
\psfig{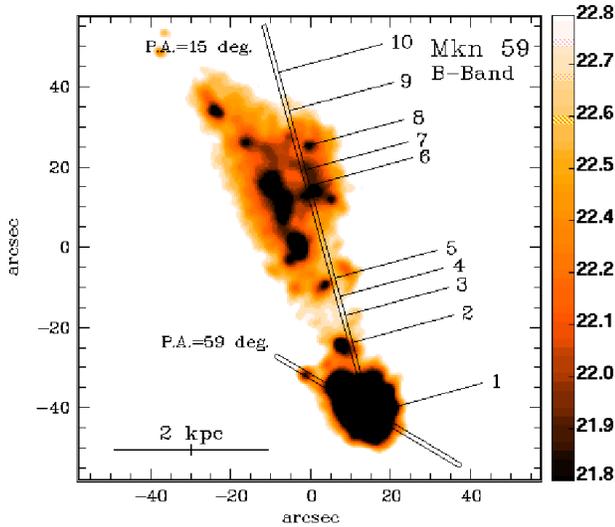}
}
\caption{$B$ band image of \object{Mkn\,59}; the two different slit positions and the regions 
investigated individually (see Section\ \ref{spectr_obs}) are shown. The grey scale to the right
displays the surface brightness levels in B mag\ arcsec$^{-2}$.}
\label{mk59_bslit}
\end{figure}

Spectrophotometric observations of \object{Mkn\,59} and \object{Mkn\,71} were obtained
on March 17th and 18th 1994, using the Ritchey--Chr\'etien spectrograph of the
KPNO 4m Mayall telescope.
We used a 2\arcsec$\times$300\arcsec\ slit with the KPC-10A grating
(316 lines mm$^{-1}$) in first order, with a GG 385 order separation filter
which cuts off all second order contamination for wavelengths blueward of
7400\AA. This instrumental setup gave a spatial scale along the
slit of 0\farcs69 pixel$^{-1}$, a scale perpendicular to the slit of 2.7
\AA\ pixel$^{-1}$, a spectral range of 3500--7500 \AA, 
and a spectral resolution of $\sim$  5 \AA. The seeing was 1\farcs5 FWHM.
\object{Mkn\,59} was observed at two slit positions 
(cf. Figure \ref{mk59_bslit}), both centered at the brightest star--forming region 
with P.A. = 59$^\circ$ and P.A. = 15$^\circ$ (close to the direction of the 
major axis). 
The total exposure time for the first orientation of the slit 
was 70 minutes and was broken up into four subexposures. 
At the second slit orientation a single 15 minutes exposure was taken.\smallskip 

% ---------- Figure6 -----------------------------------
%-----------------------------------------------------------
%
\object{Mkn\,71} was observed at a slit orientation of P.A. = 77$^\circ$ 
(see Figure\ \ref{bi_maps}), centered on the brightest star--forming region 
(\object{NGC\,2363\,I}). The total exposure time was 29 minutes, broken up 
into five subexposures. 
Three Kitt Peak IRS spectroscopic standard stars were observed during each 
night for flux calibration. The airmasses during all observations were 
$\leq$ 1.25, therefore a correction for atmospheric dispersion was not
necessary. Spectra of He-Ne-Ar comparison lamps were obtained before and after 
each observation to calibrate the wavelength scale.\smallskip 

\begin{figure*}
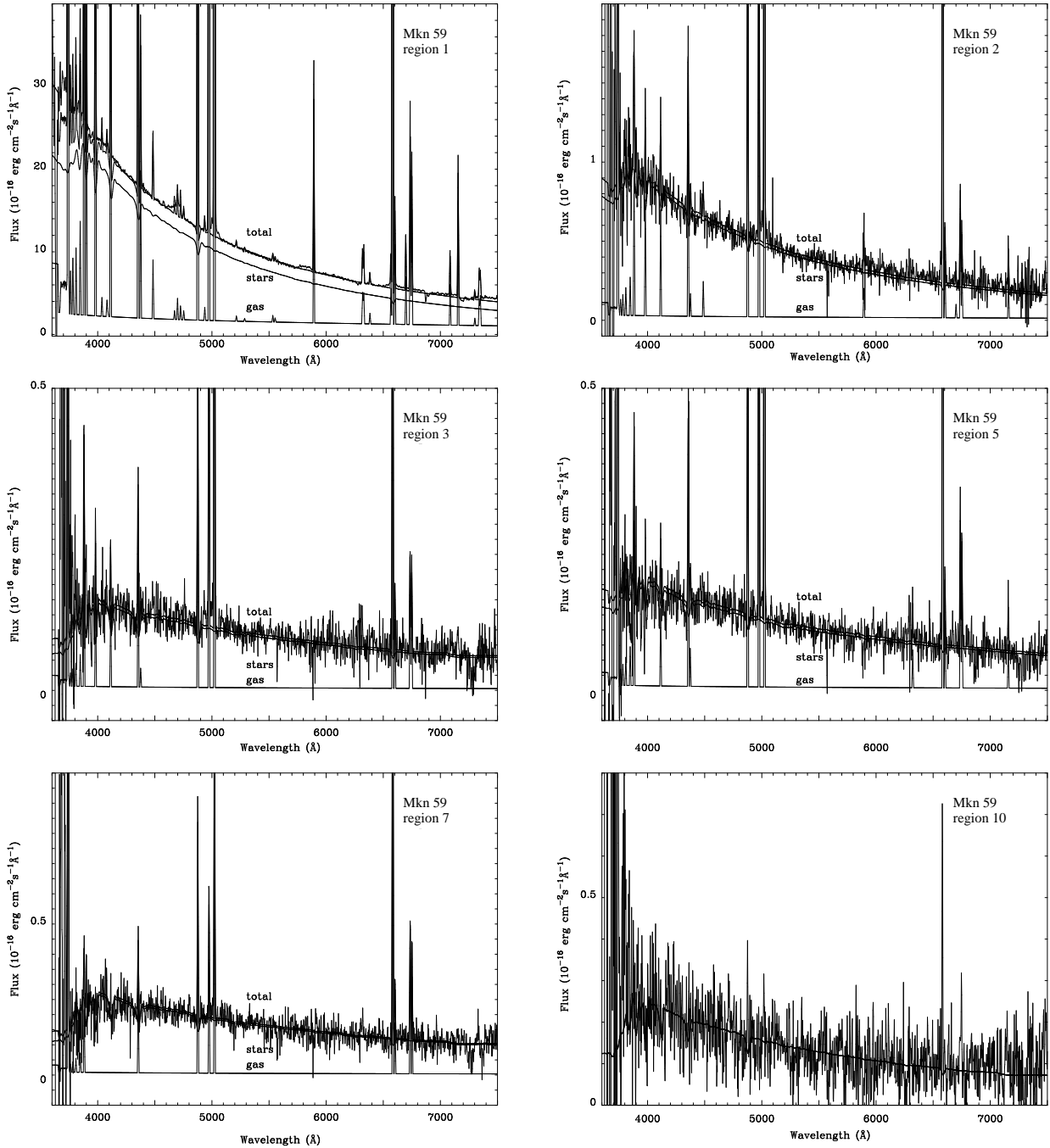

\centerline{\psfig{file=9237.f12,height=6cm,angle=270}\hspace*{8mm} \psfig{file=9237.f13,height=6cm,angle=270}}
\centerline{\psfig{file=9237.f14,height=6cm,angle=270}\hspace*{8mm} \psfig{file=9237.f15,height=6cm,angle=270}}
\centerline{\psfig{file=9237.f16,height=6cm,angle=270}\hspace*{8mm} \psfig{file=9237.f17,height=6cm,angle=270}}
\caption[]{\label{Fig1}
Spectra of different regions of \object{Mkn\,59} extracted along the major axis slit at P.A. = 
15$^\circ$ (thin lines). Overplotted are spectral energy distributions (SEDs) of model composite stellar populations
and gaseous SEDs. The SEDs of regions No.\ 3,
7 and 10 are redder than those of the other regions and refer to the underlying stellar 
component. It is obvious that the contribution of ionized gas to the observed SED is 
much larger in region 1 than in other regions further to the northeastern direction of the main body of
the BCD.}
%is significant  The contribution of the ionized gas is large in region 1, compared to the other regions.}
\end{figure*}

\begin{figure*}
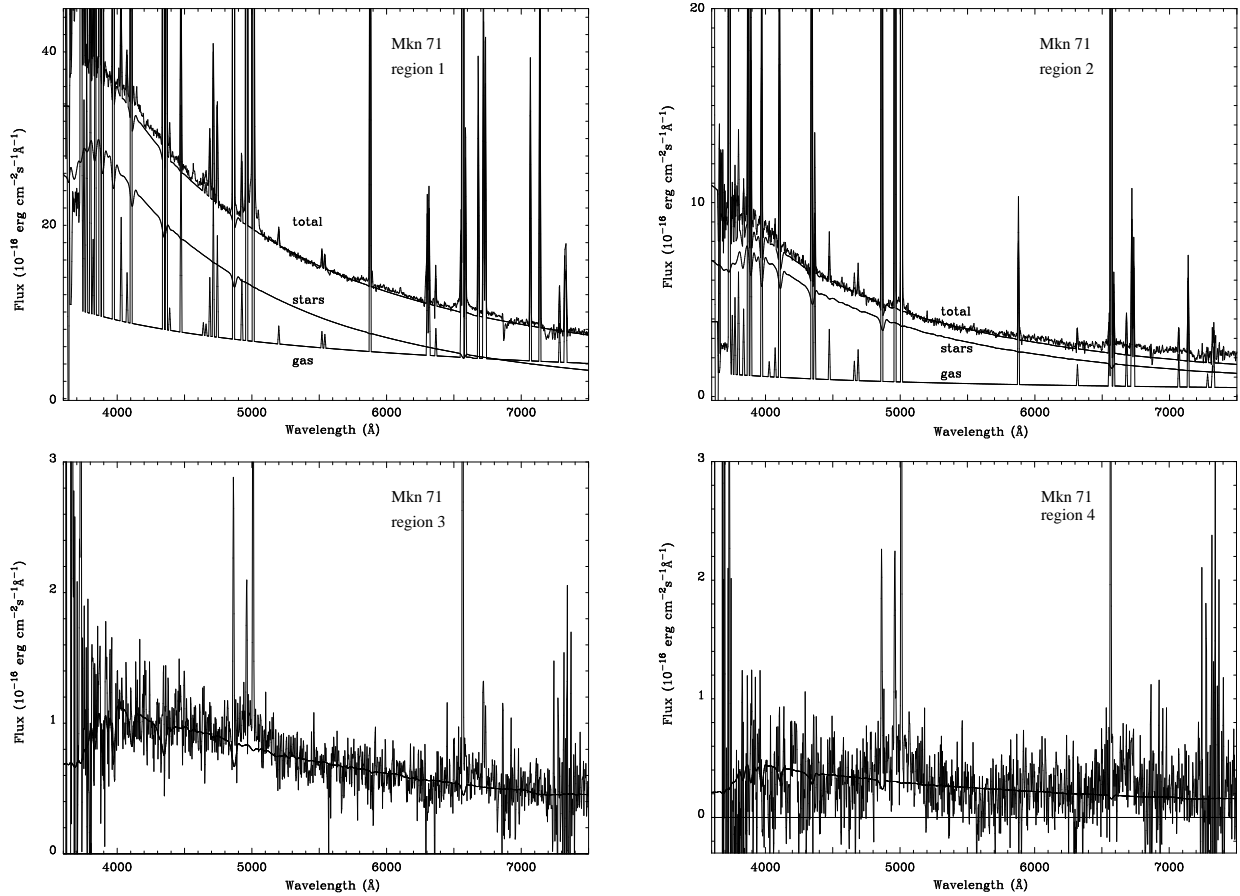

\centerline{\psfig{file=9237.f18,height=6cm,angle=270}\hspace*{8mm} \psfig{file=9237.f19,height=6cm,angle=270}}
\centerline{\psfig{file=9237.f20,height=6cm,angle=270}\hspace*{8mm} \psfig{file=9237.f21,height=6cm,angle=270}}
\caption[]{\label{Fig2}
Spectra of the regions in \object{Mkn\,71}, extracted from the slit at P.A. = 
77$^\circ$ (thin lines). Overplotted are spectral energy distributions (SEDs) of model composite stellar populations and gaseous SEDs. The SED of the region No.\ 4
is redder than those of other regions and matches the underlying stellar 
component. Note the large contribution of the ionized gas in region 1.}
\end{figure*}

The data reduction was done with the IRAF software package. 
The two--dimensional spectra were bias--subtracted and flat--field corrected.
Then the IDENTIFY, REIDENTIFY, FITCOORD, TRANSFORM, BACKGROUND and CALIBRATE 
routines were used to perform the wavelength calibration, correction for distortion
and tilt, night sky subtraction and flux calibration for each frame. 
One--dimensional spectra were extracted from the flux--calibrated two--dimensional 
spectra using the IRAF task APALL. 
For \object{Mkn\,59} we chose ten 7\arcsec$\times$2\arcsec\ regions along the major axis
of the galaxy (i.e the slit at P.A. = 15$^{\circ}$), the locations of which are 
indicated in Figure \ref{mk59_bslit}. Six of these spectra are shown in Figure \ref{Fig1}.
From the slit at P.A. = 59$^{\circ}$ the subspectra were extracted at each
pixel row along the slit, so that the resulting areas were 0\farcs69$\times$2\arcsec\ each.
For \object{Mkn\,71} (see Figure\ \ref{bi_maps}), the subspectra from regions 1 and 2 were 
extracted from 7\arcsec$\times$2\arcsec  regions, while for regions 3 and 4 areas of 
14\arcsec$\times$2\arcsec\ and 28\arcsec$\times$2\arcsec\ were used to obtain a sufficiently 
high ${S/N}$. The extracted spectra are shown in Figure \ref{Fig2}.
\smallskip

The emission line intensities were measured utilizing a Gaussian profile fitting. 
All spectra were corrected for interstellar extinction, where the extinction coefficient 
$C$(H$\beta$) was derived from the hydrogen Balmer 
decrement using the equations given in Izotov et al. (\cite{itl94}) and the theoretical
hydrogen emission line flux ratios from Brocklehurst (\cite{brocklehurst71}).
The variations of $C$(H$\beta$) along different regions (Tables \ref{Tab1} and 
\ref{Tab1a}) arise mainly from observational uncertainties (imperfect focussing 
at the blue and red ends of the spectra), 
and from different combinations of Balmer lines that could be reliably 
measured in each subspectrum to determine the local extinction.
While each spectrum was corrected using its individual value of $C$(H$\beta$), 
images were extinction--corrected adopting a uniform value of
$C$(H$\beta$)=0.09 and $C$(H$\beta$)=0.12 for \object{Mkn\,59} and 
\object{Mkn\,71}, respectively (Guseva et al. \cite{guseva98b}).

The electron temperature $T_e$(\ion{O}{iii}) was derived from the 
observed flux ratio [\ion{O}{iii}]($\lambda$4959+$\lambda$5007)/$\lambda$4363. 
In cases where the [\ion{O}{iii}] $\lambda$4363 emission line was not detected, 
we used the so--called {\it upper branch} of the Edmunds \& Pagel (\cite{edmunds84}) 
calibration of the total oxygen emission line flux 
[\ion{O}{ii}]$\lambda$3727+[\ion{O}{iii}]($\lambda$4959 + $\lambda$5007) 
vs. the electron temperature to determine $T_e$ and derived the oxygen abundance 
following van Zee et al. (\cite{vanzee98b}).

The observed ($F$($\lambda$)) and corrected ($I$($\lambda$)) emission line
fluxes relative to the H$\beta$ emission line fluxes for 3 regions
in \object{Mkn\,59} (slit P.A. = 15$^\circ$) are listed in Tables \ref{Tab1} and \ref{Tab1a}. 
The tables contain only the regions where the emission line [\ion{O}{iii}]$\lambda$4363 
was detected at a $S/N$ that allowed for a reliable flux measurement. Also listed 
are the extinction coefficient $C$(H$\beta$), the observed flux
of the H$\beta$ emission line and its equivalent width $EW$(H$\beta$) along with
the equivalent width of hydrogen absorption lines $EW$(abs). 
The line intensities of the brightest regions (1 and 2) of \object{Mkn\,71} are 
presented in Izotov et al. (\cite{itl97}). 

Applying the electron temperature $T_e$(\ion{O}{iii}), ionic abundances of O$^{2+}$, 
Ne$^{2+}$ and Ar$^{3+}$ were derived.
The temperature $T_e$(\ion{O}{ii}) was inferred, according to Izotov et al. (\cite{itl94}, 
\cite{izotov96}), from the relation between $T_e$(\ion{O}{ii}) and $T_e$(\ion{O}{iii}) using 
\ion{H}{ii} region photoionization models by Stasi\'nska (\cite{stasinska90}).
From $T_e$(\ion{O}{ii}), the O$^+$ and N$^+$ ionic abundances were determined, while the 
intermediate value of the electron temperature $T_e$(\ion{S}{iii}) served to derive the ionic 
abundances of Ar$^{2+}$ and S$^{2+}$ (Garnett \cite{garnett92}).
The [\ion{S}{ii}]$\lambda$6717/$\lambda$6731 ratio was used to determine the electron 
number density $N_e$(\ion{S}{ii}). Total heavy element abundances
were derived after correcting for unseen stages of ionization following
Izotov et al. (\cite{itl94}, \cite{itl97}) and Thuan et al. (\cite{til95}).

The resulting ionic and heavy element abundances for the 3 \ion{H}{ii} regions in \object{Mkn\,59} 
with measured [\ion{O}{iii}]$\lambda$4363 are given in Table \ref{Tab2} along with 
the adopted ionization correction factors (ICF) while the heavy element abundances
in the two brightest knots of \object{Mkn\,71} are given in Izotov et al. (\cite{itl97}).

%*********************************
%   Table 2
%*********************************
\begin{table*}[t]
\caption{Heavy Element Abundances in \object{Mkn\,59}}
\label{Tab2}
\begin{tabular}{lccc} \\ \hline
Property&region 1&region 2&
region 3\\ \hline
$T_e$(\ion{O}{iii})(K)                 	   &13,469$\pm$38      &13,069$\pm$3,299   &13,558$\pm$2,910    \\ 
$T_e$(\ion{O}{ii})(K)                      &12,979$\pm$34      &12,762$\pm$3,008   &13,027$\pm$2,626    \\ 
$T_e$(\ion{S}{iii})(K)                     &13,224$\pm$32      &12,916$\pm$2,738   &13,293$\pm$2,415    \\ 
$N_e$(\ion{S}{ii})(cm$^{-3}$)              &    79$\pm$10      &    10$\pm$10      &   536$\pm$561      \\  \\ 
O$^+$/H$^+$($\times$10$^5$)         	   &1.502$\pm$0.012    &3.001$\pm$2.164    &4.239$\pm$2.571     \\ 
O$^{++}$/H$^+$($\times$10$^5$)      	   &8.757$\pm$0.069    &6.152$\pm$4.365    &5.095$\pm$3.013     \\ 
O$^{+3}$/H$^+$($\times$10$^5$)      	   &  ...              &  ...              &  ...               \\ 
O/H($\times$10$^5$)                 	   &10.26$\pm$0.070    &9.154$\pm$4.872    &9.334$\pm$3.961     \\ 
12 + log(O/H)                       	   &8.011$\pm$0.003    &7.962$\pm$0.231    &7.970$\pm$0.184     \\ 
$[$O/H$]$$^a$              		   &--0.919$\pm$0.003~~&--0.968$\pm$0.231~~&--0.960$\pm$0.184~~~\\  \\ 
N$^+$/H$^+$($\times$10$^7$)         	   &4.902$\pm$0.040    &14.26$\pm$10.28    &19.98$\pm$12.12     \\ 
ICF(N)                              	   &6.83\,~~~~~~~~~~   &3.05\,~~~~~~~~~~   &2.20\,~~~~~~~~~~    \\  
log(N/O)                            	   &--1.486$\pm$0.006  &--1.323$\pm$0.550~~&--1.327$\pm$0.455~~~\\  \\ 
Ne$^{++}$/H$^+$($\times$10$^5$)     	   &1.639$\pm$0.014    &1.243$\pm$0.931    &1.646$\pm$1.016     \\ 
ICF(Ne)                             	   &1.17\,~~~~~~~~~~   &1.49\,~~~~~~~~~~   &1.83\,~~~~~~~~~~     \\  
log(Ne/O)                           	   &--0.728$\pm$0.006~~&--0.695$\pm$0.461~~&--0.491$\pm$0.374~~~\\  \\ 
S$^+$/H$^+$($\times$10$^7$)         	   &2.164$\pm$0.013    &4.634$\pm$1.900    &7.776$\pm$2.782     \\ 
S$^{++}$/H$^+$($\times$10$^7$)      	   &14.85$\pm$0.281    &  ...              &  ...               \\ 
ICF(S)                              	   &1.89\,~~~~~~~~~~   &  ...              &  ...               \\  
log(S/O)                            	   &--1.504$\pm$0.006~~&  ...              &  ...               \\   \\ 
Ar$^{++}$/H$^+$($\times$10$^7$)     	   &3.497$\pm$0.023    &4.652$\pm$1.738    &  ...               \\ 
Ar$^{+3}$/H$^+$($\times$10$^7$)     	   &1.049$\pm$0.080    &  ...              &  ...               \\ 
ICF(Ar)                             	   &1.02\,~~~~~~~~~~   &1.53\,~~~~~~~~~~   &  ...                \\ 
log(Ar/O)                           	   &--2.346$\pm$0.008~~&--2.108$\pm$0.266~~&  ...              ~\\  \\ 
Fe$^{++}$/H$^+$($\times$10$^7$)     	   &1.505$\pm$0.123    &  ...              &  ...               \\ 
ICF(Fe)                             	   &8.54\,~~~~~~~~~~   &  ...              &  ...                \\  
log(Fe/O)                           	   &--1.902$\pm$0.012~~&  ...              &  ...               \\  \hline
\end{tabular}

\vspace*{1ex}
$^a${[X] = logX -- logX$_\odot$. Solar abundances are from
Anders \& Grevesse (\cite{anders89}).}
\end{table*}
%

% spectral energy distribution in Mkn 59
%
% ---------- Figure7 ---------------------------------------
%
%
% spectral energy distribution in Mkn 71
%
% ---------- Figure8 -------------------------------------------
%
%
% ++++++++++++++++++++++++++++++++++++++++++++++++++++++++++++++++++++
\section {Results}
% ++++++++++++++++ 
\subsection{Structural properties of the host galaxies}
\label{res_struct}
%
% ---------- Figure9-----------------------------------------
\begin{figure*}
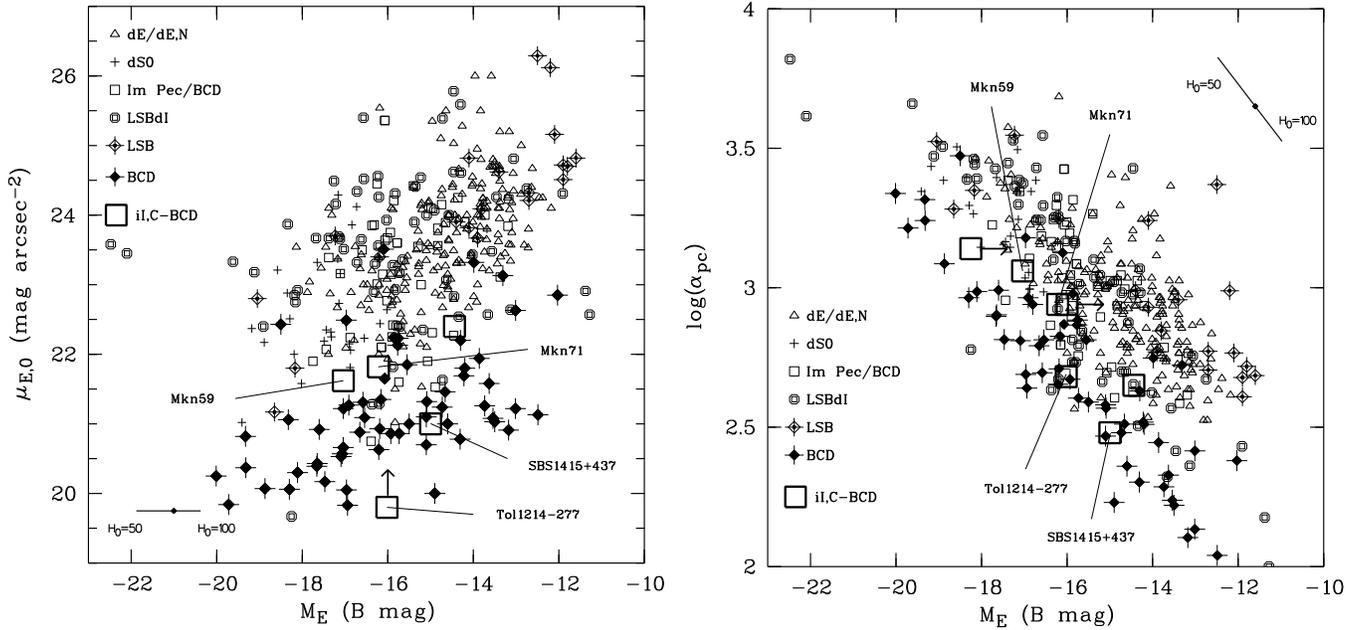

\resizebox{\hsize}{!}{
\psfig{figure=9237.f22,height=12cm,clip=,angle=270}
\psfig{figure=9237.f23,height=12cm,clip=,angle=270}
}
\caption{
$B$ band structural properties of the underlying exponential components for different types of dwarf 
and low surface brightness (LSB) galaxies, compiled in Papaderos (\cite{papaderosphd}). 
The positions of the cometary BCDs \object{Mkn\,59}, \object{Mkn\,71}, \object{Mkn\,1328}, 
\object{UM\,133}, \object{UM\,417}, \object{SBS\,1415+437} (Thuan et al. \cite{ti99}) and 
\object{Tol\,1214--277} (Fricke et al. \cite{fricke99}; cf. Section \ref{res_struct}, this work) are
indicated by big squares.
{\bf left} Extrapolated central $B$ surface brightness $\mu _{E,0}$ vs. absolute $B$ magnitude 
$M_{E}$ of the LSB components. 
{\bf right} Exponential scale length $\alpha $ vs. $M_{E}$.
Boxes with arrows to the right mark \object{UM\,133} (Telles \& Terlevich \cite{telles97}) 
and \object{Mkn\,1328} (James \cite{james94}) 
and give upper limits for $M_{E}$. 
Data points for iE/nE BCDs and other classes of dwarf galaxies are 
taken from Drinkwater \& Hardy (\cite{drinkwater91}),
\cite{papaderos96a}, Papaderos (\cite{papaderosphd}), Marlowe et al. (\cite{marlowe97}), Noeske (\cite{noeske99}), 
Noeske et al. (\cite{noeske99b}) and Binggeli \& Cameron (\cite{binggeli91}, \cite{binggeli93}), 
Caldwell \& Bothun (\cite{caldwell87}), Bothun et al. (\cite{bothun91}), 
Patterson \& Thuan (\cite{patterson96}), Vigroux et al. (\cite{vigroux86}), Hopp \& Schulte--Ladbeck (\cite{hopp91}) and Carignan \& Beaulieu (\cite{carignan89}), respectively.
\label{struct}
}
\end{figure*}
%------------------------------------------------------------
%
Recent studies have established that the underlying host galaxy of iE/nE BCDs does 
systematically differ with respect to its central surface brightness $\mu_{E,0}$ and exponential 
scale length $\alpha$ from other classes of dwarf galaxies such as dIs and dEs 
(\cite{papaderos96b}, Patterson \& Thuan \cite{patterson96}, Marlowe et al. \cite{marlowe97},
Papaderos \cite{papaderosphd}, Salzer \& Norton \cite{salzer98}, Marlowe et al. \cite{marlowe99}). 
This structural dichotomy is evident from Figure\ \ref{struct} (adopted from Papaderos 
\cite{papaderosphd}) showing that the central surface brightness and exponential scale length of 
the LSB component of a BCD with an absolute $B$ magnitude $M_{E}$$\sim$--16 mag are respectively 
$\ga$1.5 mag brighter and a factor of $\sim$ 2 smaller than in a typical dI/dE of equal luminosity. 
Although there is no sharp limit, a gap around $\mu _{E,0}\approx $22 $B$\,mag$/ \sq \arcsec$ 
separates the host galaxies of iE/nE BCDs from other classes of dwarf galaxies.
The same diagram shows that more luminous BCDs (i.e. systems with a host galaxy brighter than 
--16 mag in the $B$ band) follow the same trend populating systematically different areas in the 
$\mu _{E,0}$--$M_{E}$ and log($\alpha _{E}$)--$M_{E}$--planes than dIs and dEs. 
In Figure\ \ref{struct} (left panel)  we show with open squares the positions of the LSB 
components of the iI,C BCDs studied here (values derived in the $B$ band) along with the ones of 
the iI,C BCDs \object{SBS\,1415+437} (Thuan et al. \cite{ti99}), \object{Tol\,1214--277} 
(Fricke et al. \cite{fricke99}) and \object{UM\,417} (Cair\'os et al. \cite{cairos99}).   
In the right panel of the same figure we include two further iI,C BCDs for which measurements of 
the exponential scale length of their LSB components are available, \object{UM\,133} 
(Telles \& Terlevich \cite{telles97}) and \object{Mkn\,1328} (James \cite{james94}).
It may be seen that the iI,C BCDs fit into the gap between typical iE/nE BCDs and dIs/dEs 
except for \object{SBS\,1415+437} and \object{Tol\,1214--277}
which are indistinguishable from typical BCDs in the $\mu _{E,0}$--$M_{E}$ and 
log($\alpha _{E}$)--$M_{E}$ parameter space.

The uncertain distance to \object{Mkn\,59} (cf. Section \ref{intro}) does not affect the 
results stated above. As can be seen in both panels of Figure \ref{struct}, the vector 
illustrating a shift to the data points due to a change of the Hubble constant is approximately 
parallel to the sequences of data points for iE/nE BCDs and other types of dwarf galaxies. 
For a different distance, \object{Mkn\,59} therefore remains within the
gap between compact BCDs and dIs/dEs.

% ===========================================================
\subsection{Heavy element and line intensity distributions}
% ===========================================================
\label{res_spec}
%
%
% intensity distribution in Mkn 59
%
% ---------- Figure10----------------------------------
\begin{figure*}
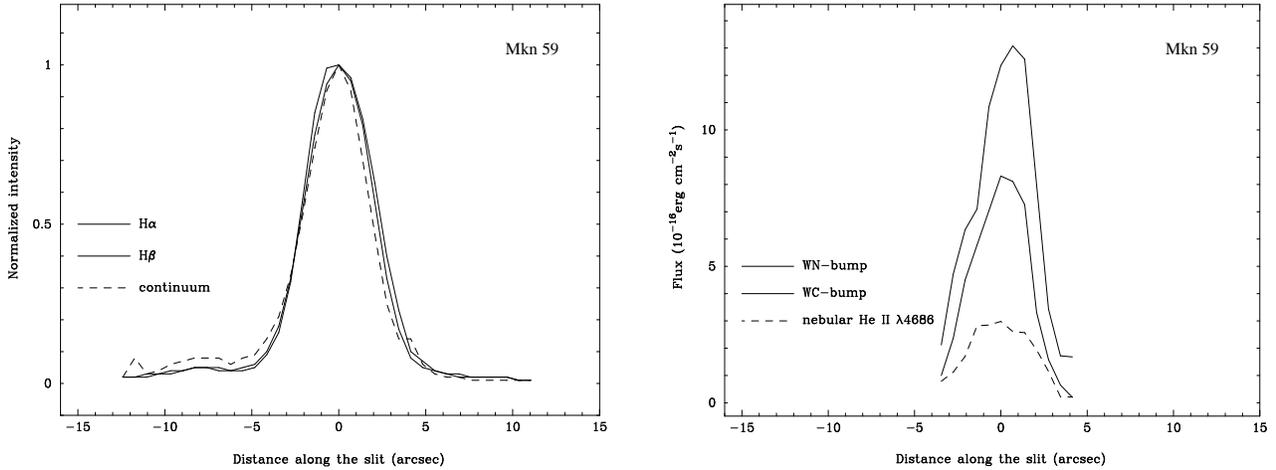

\centerline{
\psfig{figure=9237.f24,height=6.2cm,clip=,angle=270}\hspace*{7mm}
\psfig{figure=9237.f25,height=6.2cm,clip=,angle=270}
}
\caption[]{
\label{mk59_ints} 
{\bf left} 
Spatial distribution of the continuum flux near H$\beta$ (dashed line) 
and of the fluxes of the H$\beta$ and H$\alpha$ emission lines (thin and thick solid
lines, respectively) in \object{Mkn\,59} in the direction of 
P.A. = 59$^\circ$. The fluxes are normalized to the values at the brightest pixel.
{\bf right}
Spatial distribution of the WR broad bump fluxes at $\lambda$4650 (WN bump, thick 
line), $\lambda$5808 (WC bump, thin line) and the nebular \ion{He}{ii} $\lambda$4686 
emission line flux (dashed line) in Mkn 59 along the P.A. =  
59$^\circ$ direction. Note that the maximum of the WN bump appears to be
slightly shifted relative to the maximum of the H$\beta$ 
emission line, continuum and nebular flux distribution.}
\end{figure*}
%
%
% Abundances distribution in Mkn 59
%
% ----------   Figure11   -------------------------------
%
\begin{figure*}
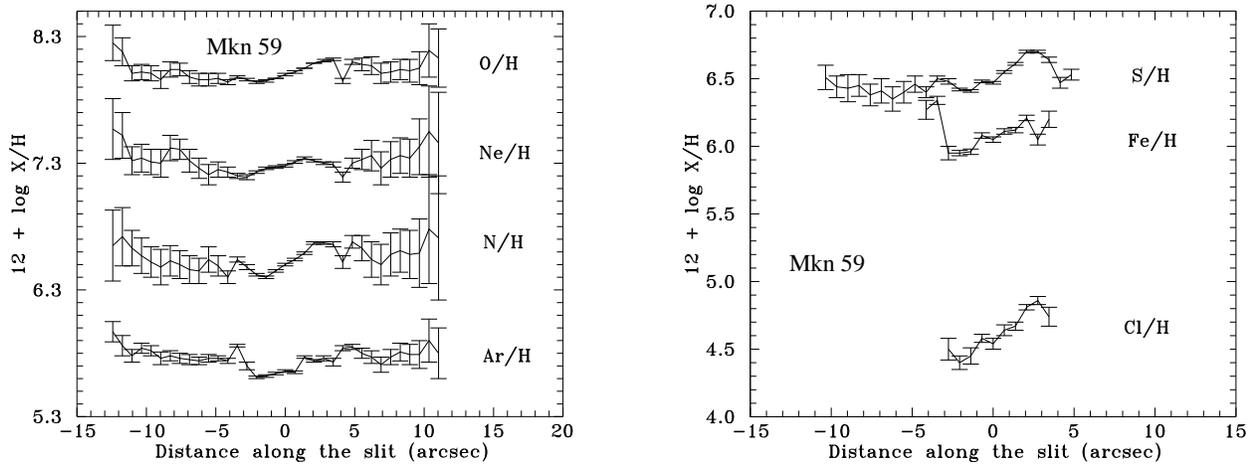

\centerline{
\psfig{figure=9237.f26,height=6.2cm,angle=270,clip=}\hspace{10mm}
\psfig{figure=9237.f27,height=6.2cm,angle=270,clip=}}
\caption[]{
\label{mk59_chem} 
Spatial distributions of the heavy element abundances in \object{Mkn\,59} along the
slit at P.A. = 59$^\circ$. }
\end{figure*}
%--------------------------------------------------------
%
In Figure \ref{mk59_ints} we show the spatial intensity distributions 
of the H$\alpha$ and  H$\beta$ lines along with that of the continuum 
adjacent to H$\beta$, as determined from the one--dimensional 
spectra extracted along the slit at P.A. = 59$^\circ$ of \object{Mkn\,59}
(cf. Figure 6). It is obvious that all three distributions are spatially coincident.
This is also the case for other nebular emission lines, 
including \ion{He}{i}, [\ion{N}{ii}], [\ion{O}{ii}], [\ion{O}{iii}] 
and [\ion{Ne}{iii}], which show the same spatial distribution as 
the H$\alpha$ and H$\beta$ emission lines.\\ 

Wolf--Rayet stars were detected in region 
No. 1 of both, \object{Mkn\,59} and \object{Mkn\,71}.
For \object{Mkn\,71}, 5 late nitrogen WR stars (WN stars) and 2 early 
carbon WR stars (WC stars) stars were found. 
The number of WC stars in \object{Mkn\,59} was determined 
to be 13 and that of WN stars to be 40 
(Guseva et al. \cite{guseva98b}, values adapted to the distances used
here). 
The spatial flux distributions of the blue $\lambda$4650 and 
red $\lambda$5808 WR bumps are shown in the right panel of
Figure \ref{mk59_ints} along with the one of the nebular 
\ion{He}{ii} $\lambda$4686 emission line. 
The red bump results primarily from WC stars, the blue bump from WN stars. 
We remark that the maxima of the H$\beta$ and WR emission are shifted by 1 pixel 
(0\farcs\ 69) along the slit at P.A. = 59$^\circ$ suggesting that the locations of 
O and WR stars may slightly differ. 
 
In all one--dimensional spectra of \object{Mkn\,59} at P.A. = 59$^\circ$ 
the [O III] $\lambda$4363 emission line is detected which allows 
a reliable determination of the electron temperatures and element abundances. 
Despite the large spatial variations of the oxygen line intensities and their 
ratios, the oxygen abundance is practically constant over the starburst region 
within the errors (Figure \ref{mk59_chem}). 
The abundances of other heavy elements (neon, nitrogen, 
argon, sulphur, iron and chlorine) show the same, nearly constant spatial 
distribution along the slit with P.A. = 59$^\circ$ through the center
of the giant \ion{H}{ii} region.
In spite of the small abundance variations within the central part, in most cases 
$\le$ 0.2 dex, we remark that a weak gradient is present for the abundances
of all heavy elements along the slit from --2\arcsec\ to 
+3\arcsec, i.e. on scales of $\sim $ 260 pc.
This suggests a local heavy element enrichment and may be related 
to the possible displacement of WR stars relative to O stars (see above).

In Figure \ref{Fig8} we show the spatial distribution of the oxygen abundance in
\object{Mkn\,59} along the major axis with P.A. = 15$^\circ$. Three regions where 
the electron temperature was derived from the [\ion{O}{iii}] 
$\lambda$4363/($\lambda$4959 + $\lambda$5007) flux ratio are indicated by stars, 
while open circles mark the regions where the [\ion{O}{iii}] $\lambda$4363 line was 
not detected, thus oxygen abundances were derived following van Zee et al. 
(\cite{vanzee98b}).
The latter abundances appear systematically higher compared to those 
where the electron temperature could be directly constrained utilizing 
the [\ion{O}{iii}] $\lambda$4363 line. However, the differences are still 
comparable to the intrinsic uncertainty of the order of $\pm$0.2 in log(O/H) ascribed 
by van Zee et al. (\cite{vanzee98b}) to their empirical calibration. 
Taking this fact into account, as well as 
the weakness of the [\ion{O}{iii}] $\lambda$4363 line in the outer parts of the 
galaxy, we shall consider the oxygen abundance as constant over the main body of 
\object{Mkn\,59} with an average value of 12+log(O/H) = 8.0.

The same analysis for \object{Mkn\,71} yields for the two brightest regions 
(regions 1 and 2; Figure \ref{bi_maps}, right), in which the
[\ion{O}{iii}]$\lambda$4363 line is observed, 
heavy element abundances of 12+log(O/H) = 7.79 and 7.77, respectively. 
The oxygen abundance in the regions 3 and 4 derived from the empirical 
calibration method described in van Zee et al. (\cite{vanzee98b}) is
larger by $\sim$ 0.3 dex. 

% ======================================
\subsection{Population synthesis models}
% ======================================
\label{popmodels}
For an analysis of the stellar content of the young ionizing clusters, 
we used the spectral energy distributions (SEDs) calculated by Schaerer \& Vacca 
(\cite{schaerer98}) for heavy element mass fractions and ages in the range between 
$Z$ = 0.001--0.02 and $t$ = 0.1 -- 10 Myr, respectively. For ages $t$ $\geq$ 10 Myr,
 we calculated a grid of SEDs for stellar populations with ages ranging from 10 Myr 
to 10 Gyr in time steps of $\Delta\log t[$yr$]=$0.1 and a heavy element mass fraction 
of $Z$ = 0.002, using isochrones from Bertelli et al. (\cite{bertelli94}) and the 
compilation of stellar atmosphere models from Lejeune et al. (\cite{lejeune98}). 
An initial mass function (IMF) with a Salpeter slope (2.35) and lower and upper mass limits 
of 0.6 \msun\ and 120 \msun was adopted. 

To study the age of the stellar populations in the galaxies and to compare 
the results of the spectral analysis with broad band photometric data, 
it is necessary to take into account both the stellar and ionized gaseous emission 
in the spectra. 
For this purpose, the stellar SED has to be separated from the gaseous emission 
following the procedure described by Guseva et al. (\cite{guseva98a}) and Papaderos 
et al. (\cite{papaderos98c}).
We added stellar SEDs calculated for instantaneous bursts with different ages 
(``single stellar populations''; {\em SSPs}) 
to the observed gaseous emission SED to match the total SED.
The contribution of the gaseous emission was scaled to the stellar emission by 
the ratio of the observed
equivalent width of the H$\beta$ emission line to the equivalent width of 
H$\beta$ expected for pure gaseous emission. 
To calculate the gaseous continuum SED at each region along the slit, the
observed H$\beta$ flux and the electron temperatures were derived from the
respective spectrum. The contributions of bound--free, free--free, 
and two--photon processes to the continuum emission were then calculated 
for the spectral range from 0 to 5 $\mu$m (Ferland \cite{ferland80}, Aller 
\cite{aller84}).
%
% ===============================================
\subsection{The ages of the stellar populations}
\label{ages}
% ===============================================
%
The number of O stars within the giant \ion{H}{ii} region complexes 
at the southeastern tips of \object{Mkn\,59} and \object{Mkn\,71} 
was derived from the H$\beta$ flux to $\approx$ 4740 and 2010, respectively, 
following the prescriptions by Guseva et al. (\cite{guseva98b}). 
As stated in Section \ref{res_spec}, WR stars were detected in both BCDs, which, 
together with the large number of O stars, contribute strongly to the total SED. 
Taking into account the average metallicities, the ages of the brightest \ion{H}{ii} 
regions in which WR stars were detected cannot exceed 4 to 5 Myr (e.g. Schaerer \& Vacca 1998).

In Figs. \ref{Fig1} and \ref{Fig2} the modelled stellar SEDs, the 
gaseous SEDs and their coadded spectral distributions (marked as ``total'')
are shown superposed on the observed spectra for different regions of the BCDs. 
For nearly all spectra, a good agreement is achieved between the observed and 
modelled SEDs. Only in the case of region 2 of \object{Mkn\,71} 
(see Figure \ref{bi_maps}, right and Figure \ref{Fig2}), the agreement is not as good due 
probably to an imperfectly focussed telescope.

Assuming that the observed gas emission is not caused by local SF, but
 due to photoionization or shock ionization by SF outside the observed region, 
the observed SEDs are in some cases reproducible with one SSP of intermediate age
(few 10$^8$ yr) only; mostly, however, reproducing the observed SEDs required a
superposition of SSPs with different ages and mass fractions. 
Tables \ref{Tab3} and \ref{Tab4} list the ages and relative mass fractions 
of the old and young SSPs which provide the best fits to the
observed SEDs at different slit positions.
The models imply respective ages of 4 and 2.9 Myr for the stellar population
formed in the bright \ion{H}{ii} region complexes in \object{Mkn\,59} and 
\object{Mkn\,71} (slit regions 1) while for other slit positions the age of 
the younger stellar continuum was inferred to few 10 Myr.
The ages of the old SSPs could be constrained to $\la$ 2 Gyr. 
Within the uncertainties of the methods applied, an upper limit of 
$\sim$ 3--4 Gyr is possible (see the following subsections), whereas even higher
ages appear untenable. 
One has, however, to keep in mind that present spectrophotometric dating methods
cannot definitely rule out the presence of a small fraction of even older stars,
which due to their high $M/L$ would barely contribute to the SED.
%
%*********************************
%   Table 3
%*********************************
\begin{table*}
\caption{Synthetic colours of a mixture of young and old populations in \object{Mkn\,59}}
\label{Tab3}
\begin{tabular}{llccccccc} \\ \hline
region&log $t$&relative mass&log $t$&
relative mass&$(B-V)$&$(B-R)$&$(R-I)$&$(V-K)$ \\ \hline 
\multicolumn{9}{c}{a) Stellar and gaseous emission}  \\ 
1  &6.6 &0.7 &9.0     &0.3     &0.17   &--0.01&--0.32  &0.12  \\ 
2  &6.9 &0.6 &9.3     &0.4     &0.03   &0.09 &--0.03  &0.64  \\ 
3  &8.7 &1.0 &  ...   &  ...   &0.29   &0.52 &0.28   &1.93  \\  
4  &7.4 &0.1 &9.3     &0.9     &0.21   &0.42 &0.25   &1.59  \\  
5  &7.4 &0.1 &9.3     &0.8     &0.28   &0.44 &0.16   &1.46  \\ 
6  &7.0 &0.3 &9.3     &0.7     &0.04   &0.21 &0.24   &1.39  \\ 
7  &8.7 &1.0 &  ...   &  ...   &0.28   &0.58 &0.30   &2.03  \\ 
8  &7.4 &1.0 &  ...   &  ...   &0.05   &0.24 &0.27   &1.54  \\ 
9  &7.0 &0.1 &9.3     &0.9     &0.11   &0.30 &0.28   &1.51  \\ 
10 &8.3 &1.0 &  ...   &  ...   &0.15   &0.33 &0.32   &1.90  \\  
\multicolumn{9}{c}{b) Stellar emission} \\ 
1  &6.6 &0.7 &9.0     &0.3     &--0.15  &--0.17&0.00   &0.20   \\ 
2  &6.9 &0.6 &9.3     &0.4     &--0.04  &0.02 &0.11   &0.70   \\ 
3  &8.7 &1.0 &  ...   &  ...   &0.28   &0.54 &0.39   &2.08   \\ 
4  &7.4 &0.1 &9.3     &0.9     &0.18   &0.43 &0.33   &1.70   \\ 
5  &7.2 &0.2 &9.3     &0.8     &--0.01  &0.11 &0.19   &1.22   \\ 
7  &8.7 &1.0 &  ...   &  ...   &0.28   &0.54 &0.39   &2.08   \\  \hline
\end{tabular} 
\end{table*}

%*********************************
%   Table 4
%*********************************
\begin{table*}
\caption{Synthetic colours of a mixture of young and old
populations in \object{Mkn\,71}}
\label{Tab4}
\begin{tabular}{llccccccc} \\ \hline
region&log $t$&relative mass&log $t$&
relative mass&$(B-V)$&$(B-R)$&$(R-I)$&$(V-K)$ \\ \hline
\multicolumn{9}{c}{a) Stellar and gaseous emission} \\ 
1  &6.46 &1.0 &  ...   &  ...   &0.26   &0.04   &--0.58 &0.07  \\ 
2  &6.6  &0.15&9.3     &0.85    &0.03   &0.13   &--0.14 &0.62  \\ 
3  &6.6  &0.02&9.0     &0.98    &0.30   &0.57   &0.40   &2.09  \\ 
4  &8.7  &1.0 &  ...   &  ...   &0.28   &0.54   &0.39   &2.08  \\ 
\multicolumn{9}{c}{b) Stellar emission} \\ 
1  &6.4  &1.0 &  ...   &  ...   &--0.19 &--0.25  &--0.08  &--0.42 \\ 
2  &6.6  &0.15&9.3     &0.85    &--0.07 &--0.03  &0.09   &0.67   \\ \hline
\end{tabular}  
\end{table*}
% ============================================================
\subsubsection{Uncertainties of the population synthesis models}
% ============================================================
%
The goodness of the spectral fits was found to sensitively depend on the 
mass fraction and ages of the adopted young and old SSPs.
For instance, for region 1 of \object{Mkn\,59}, increasing the age of the 
young SSP from 4 to 5 Myr yielded a satisfactory fit, but a 6 Myr 
old SSP resulted in systematic residuals between the observed and modelled 
SED. The age of the old SSPs could be constrained with an accuracy of 
$\approx$ 1 Gyr at slit regions where young stellar sources provide a minor
contribution to the light %% was not dominated by bright starburst knots 
(regions 3, 7 and 10 in \object{Mkn\,59} and region 3 in \object{Mkn\,71}).
In e.g. region 7 of \object{Mkn\,59}, old SSPs up to 2 Gyr could still reproduce 
the observed SED, while already a 3 Gyr old SSP produced obvious residuals %% ($\ga$ 5\%)
to the observed spectrum. 
A comparison of the observed and modelled $EW($H$\beta )$ yielded further constraints 
to resolve ambiguities between different SED solutions. 
The assumption of high ages for the old SSP requires an increased contribution 
by the young SSP to reproduce the blue continuum, which in turn results in 
too high $EW($H$\beta )$ as compared to the observed value.

Given that the metallicities of the stellar populations in BCDs may be lower than 
those of the ionized gas (Calzetti \cite{calzetti97}, Guseva et al. \cite{guseva98a}, 
Mas--Hesse \& Kunth \cite{mashesse99}), we computed a set of models varying the 
metallicity of the SSPs around $Z=$0.002 by 0.5 dex, the smallest stepsize our
model libraries allowed. Models computed on the assumption of a metallicity
other than $Z=$0.002 failed to adequately reproduce the observed spectrum,
suggesting that the metallicity of the ionized gas is similar to that of the
stellar population.

Alternatively, to constrain the age of the underlying galaxies, not only 
an instantaneous burst was considered, but also an extended episode of 
moderate SF at a constant rate. 
This way, the presumably complex formation histories of the galaxies
were bracketed between two limiting cases. 

In the same way as described in Section \ref{popmodels}, a grid of model SEDs was 
calculated for stellar populations that were formed in extended SF episodes which 
started and ended at different times.
The best--fitting solutions for e.g. \object{Mkn\,71} correspond to a formation episode 
of the {\em underlying galaxy} which started $\sim$ 2 Gyr ago and ended $\sim$ 10 Myr ago.
Within the uncertainties, a SF episode that began up to 3 Gyr ago is possible;
higher ages of the underlying galaxies would require a 
very recent ($<10$ Myr ago) termination of their formation episode to 
reproduce the observed SEDs. 
Such a recently ongoing SF activity in the underlying galaxy in \object{Mkn\,71} 
can be rejected from the colour--magnitude diagrams (Figure \ref{cmds}); in the left panel (a), 
obtained at the starburst region \object{NGC\,2363}, stars are absent between the actual 
burst (younger than $\sim$ 10 Myr) and the last SF episode of the underlying galaxy
which seems to have occurred more than $\sim$ 20 Myr ago (note the red
supergiants at the tip of the 25 Myr isochrone). 
Also in the right panel (b), no stars much younger than $\sim$ 20 Myr 
are present. This indicates that, in the {\em galaxy underlying the starburst}, 
no significant SF occured during the last $\ga$ 20 Myr, in agreement with 
the ground--based CMDs by Aparicio et al. (1995). 
A  similar analysis for \object{Mkn\,59} indicates, within the unvertainties, 
an upper age limit of $\sim$ 4 Gyr.

%
%**********************************************************
% Figure 8
%**********************************************************
% Oxygen distribution in Mkn 59
%
% ---------- Figure12---------------------------------
\begin{figure}
\resizebox{\hsize}{!}{
\psfig{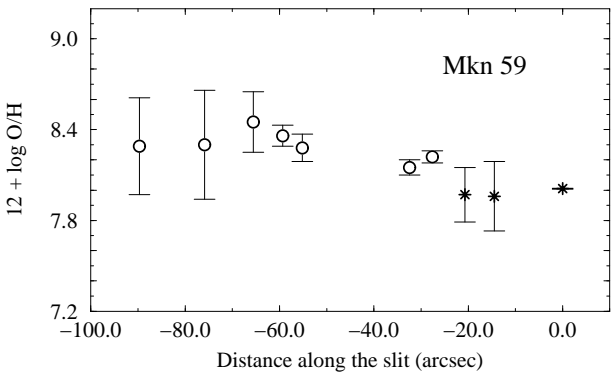}
}
\caption[]{
\label{Fig8} 
Spatial distribution of the heavy element abundances in \object{Mkn\,59} along the slit
at P.A. = 15$^\circ$. Stars mark the regions where the [\ion{O}{iii}]
$\lambda$4363 emission line was detected, open circles correspond to regions where
the oxygen abundance was derived from the empirical relation by van Zee et al. 
(\cite{vanzee98b}).}
\end{figure}
%----------------------------------------------

% ============================================================ 

\subsubsection{Consistency with imaging data; further implications}
\label{consistency}
The $EW$(H$\beta$) of region 1 in \object{Mkn\,59} and 
\object{Mkn\,71} amount to 150\AA\ (Table\ \ref{Tab1}) and 
316\AA (Izotov et al. 1997), respectively.
For these regions the broad band colours calculated from the synthetic SEDs, 
including and ignoring gaseous emission (Tables \ref{Tab3} and \ref{Tab4})
differ from each other by up to 0.5 mag. This demonstrates that a reliable 
dating of stellar populations within a starburst environment requires a  
correction for the colour shift induced by ample gaseous emission.

\begin{figure}
\resizebox{\hsize}{!}{\psfig{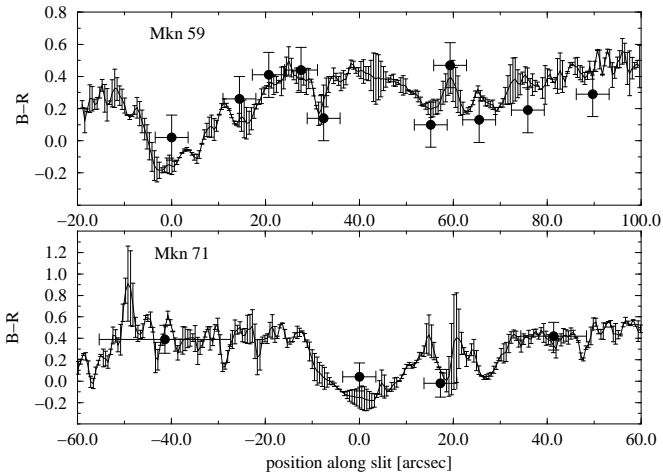}}
\caption{
Observed colour distributions along the slits extracted from colour maps, compared to 
the colours computed from population synthesis solutions (black dots) for the analyzed
subspectra.
All colours are derived on the assumption of a uniform extinction and are transformed 
to the standard Johnson--Cousins UBVRI system specified by Bessell (\cite{bessell90}), 
Bessell et al. (\cite{bessell98}).
{\bf top:} \object{Mkn\,59}, P.A. = 15\degr , {\bf bottom:} \object{Mkn\,71},
P.A. = 77\degr . 
Vertical error bars of the synthetic colours include cumulative calibration uncertainties of
the colour maps and population synthesis models, and of the transformation between the 
filter systems; horizontal error bars give the length of regions along the slit from which 
the individual subspectra were extracted. 
The alignment uncertainty between the observed and synthetic colors in direction of the 
slit (i.e. along the abscissa) is $\approx$\ 5\arcsec\ .
The error bars of the observed colour distributions give the 1$\sigma$ colour change for 
a 2\arcsec\ shift of the extracted colour map region perpendicular to the slit direction.}
\label{slitcols}
\end{figure}

The consistency of the observed broad band colours with the spectral 
population synthesis results was checked by extracting the areas covered by the 
spectrograph slits from the colour maps of \object{Mkn\,59} and \object{Mkn\,71} 
(cf. Section \ref{obred_ca}).
Figure \ref{slitcols} shows the modelled and observed colours along 
the slits after the latter were transformed to the Johnson--Cousins 
system (Bessell \cite{bessell90}).
%
% ---------- Figure13 ----------------------------------
%----------------------------------------------------------------------------
%
Generally, the observed  and synthesized colours agree well with each other 
(within 1$\sigma$ uncertainties), given that i) the synthesized colours are 
averaged values over regions covering several arcsec length along the slit, 
and ii) a minor displacement between the positions of a region in the colour map 
and of the slit may significantly alter the observed colours 
(see Figure \ref{mk59_bslit} and the 1$\sigma$--variations of the observed colours 
in Figure \ref{slitcols}), as the \ion{H}{ii} regions are typically very compact. 

Transforming the colours of the LSB component as derived from the colour profiles
(Figure \ref{profs}) to the Johnson--Cousins system yields a $B-R$ $\la$0.6 for 
\object{Mkn\,59} and \object{Mkn\,71}. This is in good agreement with the upper 
limits of the %% calibration--independent 
colours derived from the spectra (cf. Tables \ref{Tab3} and \ref{Tab4}). 
The constancy of the colour profiles for photometric 
radii where the starburst light contribution becomes negligible (i.e. for $R^*\ga P_{25}$)
indicates that in the {\em underlying galaxies} population gradients are absent. 
Therefore, the upper age limits as being derived for either BCD along the 
spectrograph slits can be considered to be valid for the entire underlying
galaxy.
%
% ================================
\section {Discussion}
% ================================
%
% ============================================================
\subsection {The metallicity distribution}
% ============================================================
\label{metal_discuss}
Compared to BCDs in general, which show a metallicity distribution that peaks around 
$\approx $ 1/10 $Z_{\sun}$ and steeply decays towards lower metallicities 
(Kunth \& Sargent \cite{kunth86}), the metal abundances of \object{Mkn\,59} 
($1/8Z_{\sun }$) and \object{Mkn\,71} ($1/14Z_{\sun }$) are not exceptional.
The nearly constant metallicity along the major and minor axes 
of \object{Mkn\,59}, as well as along the minor axis of \object{Mkn\,71} 
(Figs. \ref{mk59_chem} and \ref{Fig8}; see also Gonz\'alez--Delgado et al. 
\cite{gonzalez94}), suggests that large--scale mixing processes in the ISM of both BCDs 
were at work on time scales of a few 10$^6$ yr. 
Otherwise, one would expect measurable metallicity enhancements in the vicinity of regions 
of ongoing or recent SF. Furthermore, the transport of metal--enriched warm and hot gas 
on scales of up to $\sim$ 1 kpc from the starburst region, an observational signature of 
which is the formation of supergiant shells, is commonly seen in the ISM of star--forming 
dwarf galaxies (cf. e.g. Marlowe et al. \cite{marlowe95}, Hunter \& Gallagher \cite{hunter97}, 
Bomans et al. \cite{bomans97}, Brinks \& Walter \cite{brinks98}, Papaderos \& Fricke 
\cite{papaderos98a}, Strickland \& Stevens \cite{strickland99}; cf. also Figure \ref{mk71_hb}, 
this work).
The dilution of heavy elements in the vicinity of starburst regions may even
be powered by galactic outflows. These were predicted by numerical models 
to develop within dwarf galaxies (Vader \cite{vader86}, 
De Young \& Gallagher \cite{deyoung90}, De Young \& Heckman \cite{deyoung94}, 
Mac Low \& Ferrara \cite{maclow99}) and 
proved necessary to account for the much lower metallicities observed in BCDs with 
respect to those predicted from closed--box evolutionary synthesis models 
(Kr\"uger \cite{krueger92}, Lisenfeld \& Ferrara \cite{lisenfeld98}). 
The development of such a large--scale perturbation of the ambient 
gaseous component does not appear unreasonable given 
the high--velocity ($\sim$  10$^3$ km\ sec$^{-1}$) gaseous motions 
revealed spectroscopically by Izotov et al. (\cite{izotov96}) 
and Roy et al. (\cite{roy91}, see also Figure \ref{mk71_hb}, this work) 
in \object{Mkn\,59} and \object{Mkn\,71}, respectively.\smallskip

The small variations of the heavy element abundances along the giant \ion{H}{ii} 
region complex of \object{Mkn\,59} (Section \ref{res_spec}), as well as the possible 
spatial offset of the O-- and WR stars, may be explained
by a relocation of SF processes within a time span of few Myr. 
Propagating SF on a linear scale of $\sim$ 400 pc within the last 10 Myr has also 
been suggested by Drissen et al. (\cite{drissen99}), to account for the age differences 
of the young star clusters \object{NGC\,2363} I and II within \object{Mkn\,71}.
%
%================================================================
\subsection {Morphology vs. structural properties and age}
\label{correl_discuss}
% ================================================================
%
The two iI,C BCDs investigated here appear similar with respect to 
the age and structure of their exponentially distributed host galaxies. 
This raises the question of whether the entire class of iI,C BCDs 
shares, besides a morphological resemblance, a set of common physical properties, 
such as an intermediate age and structural properties bridging the gap between iE/nE 
BCDs and dIs/dEs. 
A literature search for data on other nearby cometary BCDs does indeed provide some
support to this hypothesis. 

{\it\object{Mkn\,1328}$/$\object{VCC\,1374}}:
NIR surface pho\-to\-metry (James \cite{james94}) yields an exponential scale length of 
$\alpha$=0.88 kpc for its LSB component. With its integrated absolute $B$-magnitude
of --16 mag, and taking into account that a burst raises the $B$ luminosity of a BCD 
by typically $\sim$ 0.75 mag (\cite{papaderos96b}, Salzer \& Norton \cite{salzer98}),
\object{Mkn\,1328} resides presumably in between the parameter spaces populated by BCDs 
and dIs in the log($\alpha _{E}$)--$M_{E}$--plane (cf. Figure 9).   
From the integral $(B\!-\!H)$ colour of $\approx$ 0.6 mag, the age of its host galaxy may 
be estimated to $\la $ 6 Gyr, following the predictions by Kr\"uger (\cite{krueger92})
 and Kr\"uger et al. (\cite{krueger95}) when a burst parameter $b<$ 0.01 is adopted.
The metallicity was determined to 1/8 $Z_{\sun}$ (Kinman \& Davidson \cite{kinman81}).

{\it\object{UM\,133}}: 
For the observed metallicity (1/17 $Z_{\sun}$, Telles \cite{telles95}),
the integral $(R-I)$ colour of 0.4 mag derived by  Telles \& Terlevich 
(\cite{telles97}) is consistent with an age of $\sim $ 4 Gyr.
With $M_{V}$=--18.25, and an exponential scale length 
of $\alpha$=1.39 kpc for its LSB component (Telles \& Terlevich \cite{telles97}), 
this object falls also into the gap between BCDs and dIs.

{\it\object{UM\,417}} ($Z$$\approx$ 1/13 $Z_{\sun}$; Campos--Aguilar at al. \cite{campos93}). 
Surface photometry of the LSB component yields an intermediately compact structure, too 
($\alpha_{E}= $ 0.45 kpc, $M_{E}= $ --14.4 $B$\ mag and $\mu _{E,0}= $ 22.4 $B$\ mag$/\sq\arcsec$; 
Cair\'os et al. \cite{cairos99}).
The same authors deduce a $(B-V)$ color for the LSB component of  $\la$ 0.5 mag, 
compatible with an age of few Gyr.

{\it\object{SBS\,1415+437}} (Thuan, Izotov \& Foltz \cite{ti99}): With a probable age
of $\la$ 100 Myr and an extremely low metallicity (1/21$Z_{\sun}$), this galaxy 
is a young galaxy candidate.
Its LSB component, however, does not fit within the structural gap between BCDs and dIs; 
with $\mu _{E,0}$=21.0 $V$\ mag$/\sq\arcsec$, $\alpha_{E}$=0.30 kpc and 
$M_{E}$=--14.95 $V$\ mag, it is rather comparable to a compact BCD. 
On the other hand, as this object apparently still undergoes its first major episode of 
SF, it is likely that its stellar LSB component has not yet been fully built 
but, as suggested by the age gradient along its major axis, delineates the trail 
along which SF has occurred. 

{\it\object{Tol\,1214--277}}:
Similar to \object{SBS\,1415+437}, this is an extremely
low--metallicity iI,C BCD ($Z\approx Z_{\sun}/$23, Terlevich et al. 
\cite{terlevich91}).
Recent surface photometry studies with the VLT (Fricke et al. \cite{fricke99}) yield
$\alpha_{E}\approx $ 0.48 kpc and $M_{E}\approx $ --16 $B$\ mag. 
The average colours of the LSB component, $(U-B)\la$ --0.4 mag and 
$(B-R)\approx$ +0.3 mag, suggest an unevolved galaxy with an age of few 10$^8$ yr.

The examples given above call for a further investigation of the following 
two hypotheses:\smallskip

\noindent (i) Cometary BCDs are relatively young objects, with ages not exceeding a few Gyr,
thus systematically younger than the majority of BCDs classified iE/nE.

\noindent (ii) Except for the candidate young extremely metal--deficient
galaxies ($\tau \sim 10^8$ yr) among them, cometary BCDs show structural 
properties of their host galaxies being typically intermediate 
between those of iE/nE BCDs and dIs/dEs.

Provided that the above hypotheses will be strengthened by an investigation 
of a larger sample of nearby iI,C BCDs, one may expect  
the iI,C morphology to occur more frequently in dwarf objects at higher redshifts.
Indeed, examples of high-$z$ galaxies displaying a comet--like morphology have  
frequently been reported (cf. e.g.  Dickinson \cite{dickinson96} for a 
galaxy cluster at $z$=1.15).
Furthermore, a few medium-redshift objects with cometary morphology 
were included in the sample of G\'uzman et al. (\cite{guzman98}, Figure 1), notably 
the Faint Blue Galaxy (FBG) \object{HERC\,13088}, at a redshift $z$=0.436.
With a total luminosity of $M_{B}$=--21.5 mag and an exponential scale length 
$\alpha\approx$ 2.8 kpc, this galaxy appears an upscaled version of the
nearby iI,C BCDs studied here. It is worth noting that the rest--frame $(B-V)$ 
colours of the LSB host of the latter FBG, $\sim$ 0.35 mag, also suggest an age of 
$\la$ 5 Gyr.

These findings provide support to the hypotheses 
proposed in Section\ 1, namely that a strong off--center burst ontop a dwarf 
LSB component as observed in iI,C BCDs is primarily not a stochastic event 
but occurs in systems considerably younger than iE/nE BCDs, probably differing  
from the latter with regard to their structural properties. 
These results emphasize the need for a detailed investigation of the 
processes leading to the development of iI,C morphology in dwarf 
galaxies. We shall briefly remark on that issue in the next section.
%

% ====================================================
\subsection{An elongated structure of high gas density?}
\label{sb_discuss}
% ====================================================

In the majority of BCDs, the centrally concentrated \ion{H}{ii} complexes 
where massive SF takes place are typically found to almost coincide
with maxima in the surface density of an extended \ion{H}{i} envelope
(Taylor et al. \cite{taylor94}, Simpson \cite{simpson95}, van Zee et al. \cite{vanzee98a}). 
Since a high gas density is necessary to sustain SF, such 
a condition is expected to be fulfilled near the end of the stellar 
body of an iI,C BCD.
Age gradients of the star--forming regions along the stellar body's major axis seem to be 
frequent attributes of such a system (Barth et al. \cite{barth94}; Aparicio et al. 
\cite{aparicio95}; Thuan et al. \cite{ti99}).  
These may be plausibly explained as due to propagation of SF activity (cf. Thuan et al. 
\cite{thuan87}) along the major axis of an elongated \ion{H}{i} body.

Given these observational results, it seems that the \ion{H}{i} halos of iI,C BCDs have 
elongated central concentrations with major axes that coincide with those of the stellar bodies.
Indeed, low--resolution \ion{H}{i} maps of \object{Mkn\,59} (Wilcots et al. \cite{wilcots96}) and 
\object{Mkn\,71} (Wevers et al. \cite{wevers86}) show the stellar components to follow prolate 
central concentrations within elongated \ion{H}{i} clouds.  
A similar distribution of the gaseous halo does not seem rare among magellanic irregulars 
(Wilcots et al. \cite{wilcots96}) and has been interferometrically mapped also in BCDs in 
different evolutionary stages (see e.g. \object{II\,Zw\,40}, van Zee et al. \cite{vanzee98a},  
and \object{I\,Zw\,18}, van Zee et al. \cite{vanzee98c}).
Moreover, the young BCD \object{SBS\,0335--052} (Izotov et al. \cite{izotov90}) is forming
within an elongated gas cloud with a probably primordial chemical composition 
(Thuan \& Izotov \cite{ti97}, Lipovetsky et al. \cite{lipovetsky99}).

It might be argued that stellar bodies and \ion{H}{i} concentrations of iI,C BCDs merely 
represent edge--on disks, as suggested by the slow rotation found for the \ion{H}{i} envelopes 
of \object{Mkn\,59} and \object{Mkn\,71} as well as for the ionized gas of 
\object{SBS\,1415+437} (Thuan et al. \cite{ti99}).
The generally exponential surface brightness profiles of their stellar LSB hosts
(cf. Figure \ref{profs}) do not serve as a proof of a disk structure (Freeman
\cite{freeman70}), since exponential SBPs are also observed for e.g. small spheroidal
systems (Binggeli et al. \cite{binggeli84}) and bars in late--type spirals 
(Elmegreen et al. \cite{elmegreen96}).
However, the assumption of edge--on disks would demand the presence of 
numerous less--inclined iI,C BCDs, i.e. relatively blue disks with a single 
starburst in their outskirts, in contradiction to morphological studies 
of dwarf galaxies.
%
% =======================================================
\subsubsection{iI,C systems vs. old iE/nE BCDs}
% =======================================================

Considering that only little data is available for BCDs with morphological types 
other than iE/nE, the construction of a tentative age--morphology sequence for gas--rich 
dwarf galaxies appears premature. 
The evolution of the different types of dwarf galaxies as well as their possible
relations among each other are still sketchy, especially the triggering and 
effects of starbursts as well as the role of the Dark Matter 
(see Thuan \cite{thuan85}, Davies \& Phillipps \cite{davies88}, \cite{papaderos96b}, 
Meurer \cite{meurer98}, Swaters \cite{swaters98}, Marlowe et al. \cite{marlowe97}, 
\cite{marlowe99}; van Zee et al. \cite{vanzee98a}, Gil de Paz et al. \cite{gildepaz99}).

However, we argue that their relatively low ages would make iI,C BCDs a possible link 
between the extremely young ($\tau\sim $ 100 Myr) dwarf galaxy candidates, most of which 
belong to the i0 class, and the evolved iE/nE BCDs, for which a {\it lower age limit} of 
2 Gyr has been invariably derived (Kr\"uger \& Fritze--v.Alvensleben \cite{krueger94}, 
Noeske \cite{noeske99}).
In the young BCDs investigated thus far, a propagation of the SF activity has been observed 
(Papaderos et al. \cite{papaderos98c}, Izotov et al. \cite{izotov99}, Thuan et al. 
\cite{ti99}). Thereby, it is conceivable that after a period of several 10$^8$ yr, these
objects will gradually develop a cometary morphology with the youngest and most 
active star--forming regions located at one end of an elongated stellar body.
%
% ================================
\section {Summary and conclusions}
% ================================
%
Aiming at a better understanding of the different morphological subclasses of 
Blue Compact Dwarf Galaxies (BCDs) in the context of the evolution of star--forming dwarf 
galaxies, we have started a study of the ``cometary'' (iI,C ) subclass of BCDs. 
Contrary to the majority of BCDs where SF is confined to the inner part of an old circular 
or elliptical stellar LSB host, iI,C BCDs exhibit intense starburst activity close to one 
end of an elongated irregular LSB component. 
This intriguing morphology prompts two hypotheses: 
(i) iI,C BCDs are in fact dwarf irregulars (dIs) observed during a major stochastic enhancement 
of their otherwise moderate SF activity and 
(ii) a set of physical properties of the gaseous and stellar components favours the ignition of 
a burst with an amplitude comparable to that typically observed in other BCDs at the outskirts 
of an older LSB component. 
  
For the two nearby cometary BCDs \object{Markarian\,59} and \object{Markarian\,71}, 
a variety of deep ground--based and HST photometric and spectrophotometric data are available.
To investigate the structural properties of the LSB component we derived surface brightness 
profiles (SBPs), corrected for systematic effects which can be important at low ${S/N}$ levels.
Photometric properties of the underlying host galaxy and the superposed starburst component were 
derived by fitting a simple decomposition scheme to the SBPs; the radial extent of the starburst 
component as derived from this profile decomposition was verified using H$\alpha$-- and colour 
maps.
The spatial distributions of heavy element abundances in the starburst regions and over the main 
stellar body were derived using long--slit spectra.
After the superposed emission by ionized gas had been semi--empirically modelled and subtracted 
from the original spectra, a population synthesis analysis was carried out at several positions 
along each slit to derive the properties of the underlying young and older stellar continuum.
For Mkn\,71 we also derived $(B-V)$ colour--magnitude diagrams from HST data.
The main findings of our analysis may be summarized as follows:\medskip

1. The azimuthally averaged intensity distribution of the underlying LSB host galaxy of 
both iI,C BCDs can be approximated by an exponential fitting law with a central surface brightness 
and a linear scale length intermediate between those typically inferred for iE/nE BCDs and 
dIs/dEs. 

2. Spectral synthesis modeling of the 
starburst region and the main body implies the presence of an older population with a most 
probable age $\la $ 2 Gyr for the underlying host galaxies of \object{Mkn\,59} and 
\object{Mkn\,71}. 
Ages up to 4 Gyr for \object{Mkn\,59} and 3 Gyr for \object{Mkn\,71} are tenable
within the model uncertainties, but higher values are unlikely.

3. The average oxygen abundances were determined to be 12+log(O/H)=8.0 (1/8 $Z_{\sun}$) for 
\object{Mkn\,59} and 12+log(O/H)=7.8 (1/14 $Z_{\sun}$) for \object{Mkn\,71}, which are typical 
among BCDs. In addition, the metallicity distribution as derived for various elements in the 
vicinity of the starburst regions and along the major axis of the LSB-body shows only small 
scatter ($\sim $ 0.2 dex), suggesting that mixing of heavy elements has been efficient.

The similarity of the two objects with respect to the ages and structural properties of their 
LSB components motivated a search for published data on other cometary BCDs. We found that five 
objects for which the required data are available share similar properties with \object{Mkn\,59} 
and \object{Mkn\,71} with respect to the results (1) -- (2). These findings suggest that the 
specific starburst morphology observed in iI,C BCDs comes along with distinct physical properties 
of their LSB host galaxies, i.e. is not attributable to stochastic processes only. 
Hypotheses which we consider worth investigating are:

\noindent (i) Cometary BCDs are relatively young objects, with ages not exceeding a few Gyr, 
thus systematically younger than ``classical'' BCDs of type iE/nE. If true, the development of 
iI,C morphology may represent a late evolutionary stage of an i0 BCD before it gradually assumes 
iE/nE--characteristics. 

\noindent (ii) The underlying host galaxies of iI,C BCDs with an age $\ga$ 1 Gyr are
moderately compact, in the sense that they show central surface brightnesses and 
exponential scale lengths intermediate between 
those typically derived for iE/nE BCDs and dIs/dEs.

The strongly extranuclear location of the starburst regions and the signatures of 
propagating star--forming activity along the main stellar body of an iI,C BCD suggest 
that the surface density distribution of cold \ion{H}{i} gas in these systems resembles the 
optical morphology; published \ion{H}{i} maps support this assumption.
Spatially resolved interferometric studies will be of major importance for 
assessing the intrinsic processes regulating SF in iI,C BCDs and 
exploring possible evolutionary links to ``classical'' iE/nE BCDs.
%------------------------------------------------------------

\begin{acknowledgements}
N.G.G. and Y.I.I. thank the Universit\"ats--Sternwarte of G\"ottingen for warm hospitality.
We acknowledge the support of Volkswagen Foundation Grant No. I/72919.
Research by K.G.N, P.P. and K.J.F. has been supported by Deutsche
Forschungsgemeinschaft (DFG) grant FR\ 325/50--1, Deutsche Agentur f\"ur 
Raumfahrtangelegenheiten (DARA) GmbH grants 50\ OR\ 9407\ 6 and 50\ OR\  9907\ 7. 
We thank the referee, Dr. R.C. Dohm--Palmer,  for helpful comments and suggestions
and C. M\"oller for providing us with a calibration spectrum.
P. Papaderos thanks K. Bischoff for his assistance during the observations at Calar Alto.
T.X. Thuan thanks the partial financial support of NSF grant AST--9616863.
\end{acknowledgements} 

%------------------------------------------------------------
{}
\end{document}